\documentstyle[11pt,aaspp4,psfig]{article}  
\received{RECEIPT DATE 1998}
\accepted{ACCEPT DATE 1998}
\journalid{VOL}{JOURNAL DATE 1998}
\articleid{START PAGE}{END PAGE}

\def\be{\begin{equation}}
\def\ee{\end{equation}}
\def\RE{R_{\rm E}}

\def\eB{\erg_{_{\rm B}}}

\def\erg{\varepsilon}
\def\lambar{\lambda\llap {--}}
\def\teq#1{$\, #1\,$}                         

\def\lsim{\lower 2pt \hbox{$\, \buildrel {\scriptstyle <}\over
         {\scriptstyle \sim}\,$}}
\newcommand\gsim{\buildrel > \over \sim}
\begin{document}
\newcommand{\figureout}[3]{\psfig{figure=#1,width=7in,angle=#2} 
    \figcaption{#3} }

\title{PARTICLE ACCELERATION ZONES ABOVE PULSAR POLAR CAPS:\\
           ELECTRON AND POSITRON PAIR FORMATION FRONTS}

\author{Alice K. Harding \& Alexander G. Muslimov\altaffilmark{1}}
\affil{Laboratory for High Energy Astrophysics \\
       NASA/Goddard Space Flight Center \\
       Greenbelt, MD 20771}


\altaffiltext{1}{NAS/NRC Senior Research Associate}

\begin{abstract}

We investigate self-consistent particle acceleration near a pulsar
polar cap (PC) by the electrostatic field due to the
effect of inertial frame dragging. Test particles gain energy from the 
electric field parallel to the open magnetic field lines and lose
energy by both curvature radiation (CR) and resonant and non-resonant 
inverse Compton scattering (ICS) with soft thermal X-rays from the
neutron star (NS) surface. Gamma-rays radiated by electrons
accelerated from the stellar surface produce pairs in the 
strong magnetic field, which screen the electric field beyond a pair 
formation front (PFF).  Some of the created positrons can be
accelerated back toward the surface and produce $\gamma$-rays and pairs that
create another PFF above the surface. We find that ICS 
photons control PFF formation near the surface, but due to the
different angles at which the electron and positron scatter the soft 
photons, positron initiated
cascades develop above the surface and screen the accelerating
electric field. Stable acceleration from the NS surface is
therefore not possible in the presence of dominant ICS 
energy losses. However, we find that stable acceleration zones may
occur at some distance above the surface, where CR 
dominates the electron and positron energy losses, and 
there is up-down symmetry between the electron and positron PFFs. We 
examine the dependence of CR-controlled acceleration zone
voltage, width and height above the surface on parameters of the
pulsar and its soft X-ray emission.  For most pulsars, we find that 
acceleration will start at a height of 0.5 - 1 stellar radii above
the NS surface.
\end{abstract}


\received{}
\revised{}
\accepted{}

\section{INTRODUCTION}

The theory of particle acceleration in pulsar magnetospheres has been under 
development for almost three decades.  Although it was well known that rotating
magnetic dipoles would induce electric fields in vacuum (Deutsch 1955), it took
several years after the discovery of pulsars to realize that the huge vacuum
fields could not in practice be available for particle acceleration.
The electric field parallel to the magnetic field is at least partly 
screened by particles supplied from the stellar surface (Goldreich \&
Julian 1969) or by electron-positron avalanches 
(Sturrock 1971).  The true accelerating voltage of a pulsar must be 
determined by departures from the corotation, or Goldreich-Julian 
charge density, that could completely screen the parallel electric
field.  Several types of models have studied pulsar acceleration due
to charge deficits at different locations in the magnetosphere.
Polar cap (PC) models consider the formation of a parallel electric field 
in the open field region near the magnetic poles, while outer gap
models consider accceleration in the outer magnetosphere, near the
null charge surface (see Mestel 1998 for the most recent and
comprehensive review of pulsar electrodynamics). 
Ruderman \& Sutherland (1975; hereafter RS75) introduced a PC model 
invoking a vacuum gap due to the trapping of ions in the 
neutron star (NS) crust.  The calculations by Jones (1985, 1986), and 
Neuhauser et al. (1986, 1987) seem to favor a low value for the work 
function (at least a factor of 10 less than it was thought earlier) in 
the NS surface with a strong magnetic field. The important implication
of this study is that the possibility of free ejection of charges 
(actually of both signs) from the NS surface can be now, 
at least theoretically, justified. In this paper, we concentrate on a 
space-charge limited flow model (implying very low work function in
the NS surface) based originally on the work of Arons \&
Scharlemann (1979; hereafter AS79), who determined the electric field 
produced by the small deparature from the Goldreich-Julian charge that 
grows above the surface due to the geometry of the open dipole
field. The electric field accelerating electrons in this model
developed along only field lines that curved toward
the rotation axis (``favorably" curved field lines), so that 
acceleration occurred over half of the PC.  The parallel field 
is shorted-out at a height above the surface where the $\gamma$-rays
from accelerated particles produce sufficient electron-positron pairs
in the strong magnetic field.  The accelerating potential is thus limited
by such a pair formation front (PFF).  

These initial calculations of electron-positron 
PFFs assumed that the primary electrons began accelerating at the 
NS surface and that curvature radiation (CR) was the only
mechanism for providing pair-producing photons (Arons 1983; hereafter A83).
In recent years, it has become clear that inverse-Compton scattering
(ICS) of soft thermal X-ray photons from the hot NS surface by the 
primary electrons is also an important mechanism above the PC.  
As well as being a significant energy loss (Kardash\"ev et al. 1984, 
Xia et al. 1985, Daugherty \& Harding 1989, Sturner 1995; hereafter
S95) and radiation 
(Sturner \& Dermer 1994) mechanism, ICS can also provide photons 
capable of producing pairs. Pulsed X-rays have been detected from 
a number of pulsars which are consistent with blackbody spectra at 
temperatures around $10^5 - 10^6$ K (Ogelman 1995).  Zhang \& Qiao
(1996) and Zhang et al. (1997; hereafter ZQLH97) first explored the 
effect of the pairs from inverse-Compton photons on the acceleration 
in a Ruderman-Sutherland
type model.  They found that ICS photons may produce a
PFF sooner (at a lower altitude) than the CR 
photons would from the same accelerating electrons.  In fact in this case, 
the electrons will stop accelerating before they can emit significant
CR.  The standard models of PC acceleration thus need substantial revision.

Another effect which has never been included in PC acceleration 
models is the formation of a lower PFF due to the positrons that are
turned around and accelerated downward from the electron PFF.
Although the number of positrons which are accelerated downward is
small compared to the number of primary electrons and even to the charge 
deficit near the upper PFF, the multiplicity of the downward cascades
is quite large
(as we will discuss in Section 3.1).  Thus, the amount of charge
produced by only a small number of downward moving positrons may be 
sufficient to establish a second PFF.  Although downward going 
cascades have been discussed in previous papers (see e.g. AS79), 
their effect on the acceleration of primaries has not been investigated.
Daugherty \& Harding (1996, hereafter DH96) qualitatively discussed 
the effects of pair cascades by returning positrons, their creation
of pairs within the acceleration zone and the need for a self-consistent
model of PC acceleration.  

In this paper, we present a detailed study of the acceleration of
primary electrons and secondary (downward-moving) positrons above
a pulsar PC, assuming space-charge limited flow (free emission)
of particles from the NS surface (see Harding \& Muslimov 1998, for a
review).  We include the general relativistic
effect of inertial frame-dragging, which induces a much larger 
electric field than that expected in the flat spacetime and is not limited to 
favorably-curved field lines (Section
\ref{sec:Ell}). This is important because, as has been concluded 
earlier in papers by Fawley, Arons \& Scharlemann (1977) and
A83, the potential drops (derived for flat space-time) are
not sufficient to account for oberved pulsar $\gamma$-rays.
Both electrons and positrons suffer energy
loss and emit photons from CR and ICS.  The treatment of ICS of both
upward and downward moving particles requires revisions from 
previous studies of only upward moving particles (Dermer 1990;
hereafter D90; S95), which are presented in Section \ref{sec:IC}. We 
then compute 
the location of both electron and positron PFFs due to one-photon pair 
production as a function of magnetic colatitude and height of the lower
PFF.  We also discuss in Section \ref{sec:pfraction} 
the fraction of positrons that are turned 
around at the upper PFF.  When the electrons are assumed to accelerate 
from the NS 
surface, we find that ICS photons produce the PFFs (Section 3.1), 
in agreement with the results of ZQLH97.
However, we also find that there is substantial difference between the 
scattering of upward-going electrons and downward-going positrons by
the same thermal X-ray photons: the electrons scatter these photons at 
angles less than $\pi/2$, while the positrons scatter the photons
head-on.  Electrons produce pairs through resonant scattering, while
positrons produce pairs by scattering above the cyclotron resonance.
The photons scattered by positrons are
therefore more energetic and produce pairs in a shorter distance.  
These pairs may screen the accelerating field up to some distance
above the surface.  We find that stable, double PFFs can form only
when CR photons produce them, i.e. at a height where 
CR losses overtake ICS losses.  We
compute the height of these stable PFFs as a function of pulsar period
and surface field strength (Section 3.1). One interesting result is that
the acceleration voltage limited by CR-controlled PFFs is only 
a function of magnetic colatitude
(i.e. geometry of the open field lines), ranging between $\sim 10^7$ 
and $3\times 10^7\, mc^2$, and is insensitive to pulsar parameters such as
period and surface value of the magnetic field strength and even to
the height of the acceleration.  The stable location of the lower PFF 
depends primarily on surface magnetic field, temperature and size of
the hot polar cap, ranging between 0.5 and 1.0 stellar radius, but is 
insensitive to period. Implications of these results for high-energy
pulsar emission are discussed in Section \ref{sec:Dis}.

\section{CALCULATION OF PAIR FORMATION FRONTS} \label{sec:PFF}

We consider a test particle approach to the determination of PFF 
locations. The test particles in this case are
electrons or positrons that gain energy through electrostatic
acceleration and lose energy through radiative losses. Photons are 
created by CR and ICS 
and destroyed by magnetic pair creation.  We assume that the electric field is
completely screened at the point where the first pair is produced.  This is a
good assumption for several reasons.  First, as discussed in more
detail below, the electric field arises due to a small imbalance
between the actual charge density and the local, rotation-induced, 
Goldreich-Julian charge density. It therefore does not require much 
additional charge to short-out this field. Second, the onset of pair 
cascading occurs very quickly (DH96), so that the number of pairs
produced per primary particle increases rapidly over small distances. 
Thus, as found in A83, the width of the PFF (the screening distance of the
electric field) is very small compared to other dimensions of the problem.
Figure 1 illustrates the geometry of the calculation.  Suppose that
the PFF results from pairs produced by $\gamma$-rays of energy 
$\varepsilon_{\rm min}$ (in units of $mc^2$) radiated by particles of
energy $\gamma_{\rm min}$.  The distance of the PFF from the starting
point of the particle acceleration at $h_0$ is then:
\be \label{Sc}
S_c = {\rm min}[S_a(\gamma_{\rm min}) + S_p(\varepsilon_{\rm min})]
\ee
where $S_a(\gamma_{\rm min})$ is the distance required to accelerate
the particle until it can radiate a photon of energy $\varepsilon_{\rm
min}$, and $S_p(\varepsilon_{\rm min})$ is the pair attenuation length of 
the photon. The acceleration distance, $S_a(\gamma_{\rm min})$, is 
determined by first integrating the equation of motion of the particle 
to determine its energy as a function of its pathlength $s$: 
\be \label{dgamma}
c{d\gamma\over ds} = {e\over mc}E_{\parallel} - \dot\gamma_{_{\rm IC}} - 
\dot\gamma_{_{\rm CR}},
\ee
where $E_{\parallel}$ is the electric field induced parallel to the
magnetic field, $\dot\gamma_{_{\rm IC}}$ and $\dot\gamma_{_{\rm CR}}$
are the loss rates for ICS and CR. Discussion of these processes will 
be given in Sections 
\ref{sec:Ell} and \ref{sec:rloss}. The pair production attenuation 
length of photons radiated by the particle through either ICS or CR 
is then determined. This attenuation length $S_p(\varepsilon)$, defined to be the path length over which the 
optical depth is unity, is given by
\be \label{tau}
   \tau(\varepsilon)\; =\;\int_0^{S_p(\varepsilon)} 
T_{\rm pp}(\theta_{\rm kB},\, \varepsilon )\, ds\; =\; 1\quad ,       
\ee
where \teq{ds} is the pathlength differential along the photon momentum
vector \teq{{\bf k}}, $T_{\rm pp}$ is the attenuation coefficient for 
one-photon pair production and $\theta_{\rm kB}$ is the angle between 
\teq{{\bf k}} and the local magnetic field direction. Computation of 
$S_p(\varepsilon)$ will be discussed in Section \ref{sec:pair}.
As we will discuss in Section \ref{sec:numres}, when ICS causes the PFF$^+$
and PFF$^-$, $S_c^+$ is smaller than $S_c^-$, due to different modes
of ICS.  The PFF$^+$ will then not coincide with the start of the electron
acceleration at $h_0$.

\subsection{Polar Cap Electrodynamics} \label{sec:Ell}

We use the electric field due to inertial frame dragging above the
NS surface, first calculated by Muslimov \& Tsygan (1990,
1992; hereafter MT90, 92). The regime under which the generation 
of this field occurs is 
actually the same as implied in the electric field computations 
in flat spacetime by AS79, that of space-charge limited flow.  
An electric field must be present above the NS 
surface because as charges flow along open magnetic field lines, 
the corotation, or Goldreich-Julian charge density $\rho_{\rm GJ}$,
cannot be maintained. Even though $\rho = \rho_{\rm GJ}$ and therefore
$E_{\parallel} = 0$ at the surface, the curvature of the field lines causes the
area of the open field region, through which the particles flow, to
increase faster than $\rho_{\rm GJ}$, and a charge deficit grows with 
distance.  Thus, the $E_{\parallel}$ grows with height up to about one 
stellar radius above the surface, and then drops off. It is important
that for a nearly aligned NS in flat spacetime, the space-charge
density of the outflowing particles (electrons) proved to be almost
exactly compensated by the effective Goldreich-Julian charge density,
thus resulting in the substantial suppression of the electric field in
the region of open field lines. General
relativity causes a significant modification of the mechanism of the 
electric field induction, through the effect of dragging of inertial 
frames, a consequence of the distortion of spacetime by a rotating 
gravitating body. An observer near a
rotating mass experiences a force and must corotate to remain in an 
inertial frame, with an angular velocity that decreases with distance from 
the rotating mass.  The charge density above a NS surface must be 
computed in a local inertial frame that is rotating with respect both
to the NS and to an observer at infinity. Thus the
Goldreich-Julian charge density, which is the charge density required
to 
make magnetospheric particles drift in corotation with the star, will
differ from that in flat spacetime.  This charge difference enhances 
$E_{\parallel}$ over what it would be in flat spacetime, by a factor of 50
- 100 for a typical 1 s pulsar. The frame-dragging contribution to 
$E_{\parallel}$ that signifies the striking difference between the 
general relativistic and classical treatments of NS electrodynamics 
is proportional to  
$\cos\chi$ (see equation [\ref{Ell_1}] below), where $\chi$ is
the angle between the magnetic and spin axes of the pulsar.
[Obviously, there is also a frame-dragging contribution to
$E_{\parallel}$ which is proportional to $\sin\chi$, but it is
comparable to that produced in a flat spacetime limit.] Particle 
acceleration may therefore occur throughout the entire open field line 
region, with the relative contribution of the frame dragging component
to $E_{\parallel}$ being strongest for pulsars with small obliquities.

Muslimov \& Harding (1997; hereafter MH97) derived expressions for 
the $E_{\parallel}$
due to frame dragging in two limits: close enough to the NS 
surface such that $z = s/R \ll \theta_0$, where $\theta_0$ is the
PC half-angle, and far from the surface such that $z \gg
\theta_0$. These expressions were derived from solutions to Poisson's 
equations assuming boundary conditions $E_{\parallel} = 0$, as well as
the potential $\Phi = 0$ at the stellar surface and along the last open field
line.  We have adapted these solutions for use in this paper by
incorporating the screening effect of 
an upper boundary at $z_c = S_c/R$, i.e. of the pair formation
front, where $E_{\parallel}(z = z_c) = 0$. Although this approach 
is not fully self-consistent, in Section 2.1.1 we discuss how 
the screening of the electric field at the upper boundary could be
included in our calculations in a self-consistent way. In this
paper we also explore the situation where the positrons flowing back
to the PC surface initiate electron-positron cascades. 
Although this possibility has been discussed since the very
first papers on pulsar electrodynamics 
(see e.g. AS79), it has never
been addressed at an appropriate quantitative level. Here we attempt to 
approach this and some other related problems from the quantitative
point of view which, we believe, can advance our understanding of the 
PC electrodynamics and its relevance to the modeling of $\gamma
$-ray emission from pulsars.  

We note that the derivation of formulae for the electric field 
(potential) we exploit in the present analysis implies that the 
angle $\theta_0$ (magnetic colatitude of the polar field line) 
is small, which is a very good approximation for the 
region of open magnetic field lines in most pulsars. However, it is 
instructive to look at the general-relativistic expression for the 
Goldreich-Julian charge density that contains a contribution of order of 
$\theta ^2$ to the main term. This expression reads
\be
\rho _{\rm GJ} = - {{\Omega B_0}\over {2\pi c \alpha \eta ^3}} 
{{f(\eta)}\over {f(1)}} \left\{ \left[ \left( 1 - {\kappa \over {\eta
^3}} \right) - {3\over 2} H(\eta ) \theta ^2 \right] \cos \chi 
+ {3\over 2} H(\eta )\theta \sin \chi \cos \phi \right\}, 
\ee
where $\Omega $ is the NS rotation frequency, $B_0$ is 
the surface value of the magnetic field strength at the magnetic pole, 
$\alpha = (1-\epsilon/\eta )^{1/2}$ is the red-shift function, 
$\epsilon = r_{\rm g}/R$, $r_{\rm g}$ is the gravitational radius
of the NS, R is the stellar radius, $\eta =r/R$ is the 
dimensionless radial coordinate, $\kappa = \epsilon I/MR^2$, $I$ 
and $M$ are the moment of inertia and mass of the NS, 
respectively, and the functions $f(\eta )$ and $H(\eta )$ are
defined below (see equations [8], [16]). In formula (4) the functions 
$f$ and $H$ should be evaluated at $\eta =r/R$ and $\epsilon = 
r_{\rm g}/R$. 
In our derivation of the general-relativistic electrodynamic equations
we neglect the contribution to the electric field produced by 
the flaring of the magnetic field lines, which is a factor of 
$\sim \theta ^{-2} \sim 500~(P/0.1~{\rm s})$ (where $P$ is the pulsar
spin period) smaller than that produced by the frame dragging. This 
contribution results from the second-order terms like that in the 
square bracket of equation [4]. [In Section 2.4 we will refer to this
equation to clarify yet another important issue.] 
The small-angle approximation may not be accurate enough for the 
millisecond pulsars, for which $\theta \sim 0.3~(r/R)^{1/2} (P/2~{\rm
ms})^{-1/2}$. We also note that the effect of the electric 
field generation by the frame dragging does not depend on the particular 
configuration of the stellar magnetic field, simply because while 
the distribution of the real space charge in the acceleration region 
is mainly determined by the flaring of the magnetic field lines, 
the effect of frame dragging on the Goldreich-Julian charge density is 
independent of the geometry of the polar field lines. This means that
one cannot mimic the frame dragging effect simply by distorting the
polar magnetic flux tube.

\subsubsection{Rescaling of $E_{\parallel}$ with no upper PFF}

The possibility of electron-positron cascades near
the stellar surface initiated by the backflowing positrons unavoidably 
implies that the bottom of the polar magnetic flux tube should be treated as a
highly-conducting layer of an electron-positron plasma on the top of
the PC surface, rather than as
a regular NS surface made of iron or hydrogen atoms in the strong
magnetic field. In a more extreme case (see e.g. Wang et al. 1998), 
which seems to be less justified at the moment, the 
whole NS surface may be covered by a thin electron-positron 
plasma layer. Note that the presence of such a layer even on the top 
of the PC should substantially affect the electrodynamics and therefore the
acceleration of charged particles in the region of open magnetic field 
lines.  It will allow the free supply of charges into the acceleration
region, a requirement for space-charge limited flow.  Also, we shall 
discuss in Section 3, a self-consistent
regime with double PFFs that may significantly differ from the standard 
one implying a single PFF above the stellar surface.

In this Section and the next, we discuss the principal modifications one
must introduce
into the treatment of the acceleration region to incorporate the effects 
resulting from the occurrence of double PFFs. First of all, the 
appropriate distance variables in $E_{\parallel}$ should be rescaled by an
``effective" NS radius $R_{\rm E} = R + h_0$ (cf. Fig 1), 
which is now assumed to be the lower boundary of the acceleration, 
such that $E_{\parallel} (z = 0) = 0$, where $z \equiv s/R_{\rm E}$ is
the dimensionless altitude above the effective surface. 
The expressions for the electric field are thus symmetric between the 
lower and upper boundaries, in the sense that they allow the proper 
treatment of e.g. both primary electron and secondary positron
acceleration, which is an essential requirement for the theory. 

To clarify our analytic solutions (see Section 2.1.2) for the rescaled 
electric potential (accelerating electric field) produced between 
the lower and upper pair fronts, let us consider the expression for
the electric potential in terms of the difference between the actual 
and Goldreich-Julian charge densities $\rho - \rho _{\rm GJ}$. 
Because $\rho$ and $\rho_{\rm GJ}$ are just linear combinations of
$\sin\chi$ and $\cos\chi$, we can write the potential $\Phi$ in the
simple form (at distances from the stellar surface greater than the
PC size) as
\be \label{Phi}
\Phi = \alpha S (r) \left[ (\rho - \rho _{\rm GJ})
_{|_{\chi = 0}} \cos \chi + {1\over 2} (\rho - \rho _{\rm GJ})
_{|_{\chi = {\pi \over 2}}} \sin \chi \right] (1-\xi ^2),
\ee
where $S(r)\equiv \pi [r\theta (\eta )]^2$ is the cross-sectional 
area of the polar magnetic flux tube at the
radial distance $r$ and $(\rho - \rho _{\rm GJ})
_{|_{\chi = 0}}$ and $(\rho - \rho _{\rm GJ})_{|_{\chi = {\pi \over
2}}}$ are the coefficients of $\cos\chi$ and $\sin\chi$ 
in the expression for $(\rho - \rho_{\rm GJ})$ 
(see equations~[\ref{rho}] and [\ref{rhoGJ}]). It also says that the local
value of the electric potential is of order of the local value of 
the effective space charge (equal to the difference between the real 
space charge and induced Goldreich-Julian space charge) divided by the 
characteristic longitudinal length scale. 
The expression (\ref{Phi}) does not incorporate 
the effect of the screening of the electric field at the upper 
boundary (PFF). This effect is included in the corresponding
expressions we discuss later in Section \ref{sec:EPFF}. The relation 
(\ref{Phi}) gives the correct solution for the electric
potential and longitudinal component of the electric field for
altitudes  
greater than the PC size and satisfying the $\Phi = 0$ and 
$E_{\parallel }=0$ boundary conditions at the (effective) stellar
surface, where $\rho = \rho _{\rm GJ}$. This expression is very useful
for illustrating the modification of the potential and longitudinal
component of the electric field resulted from the redefining of the
position of the lower zero-electric field boundary (set e.g. by the 
lower PFF produced by the backflowing positrons). It explains the 
physical 
meaning of such a modification, which is simply a readjustment of the
true charge outflow (in the regime of self-limitation) to the 
local value of the Goldreich-Julian charge density at the effective 
surface, and allows us to better understand the corresponding exact analytic
solution.

We introduce the radial coordinate scaled by 
effective radius $R_{\rm E}$, $\eta \equiv 1 + z = 
(R_{\rm E} + s)/R_{\rm E}$, and transverse coordinate 
$\xi \equiv \theta/\theta(\eta)$, the magnetic colatitude scaled by 
the half-opening angle of the polar magnetic flux tube, 
\be
\theta(\eta) = \theta_0\left[\eta{f(1)\over f(\eta)}\right]^{1/2}
\ee
at radius $\eta$, where
\be
\theta_0 = \left[{\Omega \RE\over {cf(1)}}\right]^{1/2} 
\ee
is the PC half-angle at the effective surface $\RE$, and
\be \label{f}
f(\eta) = -3\left({\eta\over \epsilon}\right)^3\left[\ln \left( 
{1-{\epsilon\over \eta}}\right) + 
 {\epsilon\over \eta}\left( 1+{\epsilon\over 2\eta}\right) \right]
\ee
is the correction factor for the dipole component of the magnetic flux
through the magnetic hemisphere of radius $r$ in a Schwarzchild metric 
(MH97). \\
Note that in equation (5) 
\be
S(r) \equiv \pi {{\Omega }\over c}{{r^3}\over {f(\eta )}} .
\ee

The expressions for $\rho $ and $\rho _{\rm GJ}$, valid for any radius, 
can be written as 
(see e.g. MH97) 
\be  \label{rho}
\rho = -\sigma (r) \left[ (1 - \eta _{\ast }^2 \kappa ) \cos \chi + 
{3\over 2} \theta _0 H(1) \xi \sin \chi \cos \phi \right],
\ee
\be  \label{rhoGJ}
\rho _{\rm GJ} = -\sigma (r) \left[ (1 - \eta _{\ast }^2 
{\kappa \over {\eta ^3}}) \cos \chi + {3\over 2} \theta (\eta) H(\eta) 
\xi \sin \chi \cos \phi \right].
\ee
In these expressions 
\be
\sigma (r) \equiv {1\over 2} \left( {\Omega \over c} \right) ^2 
R^3 {{B_0}\over {f(\eta _{\ast })}} {1\over {\alpha S(r)}},
\ee
where $\eta _{\ast }=R/R_{\rm E}$ ($0.5\lsim \eta_{\ast } \lsim 1$, as
it will be shown below).

Thus, after substituting the above expressions for
$\rho $ and $\rho _{\rm GJ}$ into equation (5) we get the 
following formulae for the rescaled electric potential 
\begin{eqnarray}
\Phi & = & {1\over 2} \Phi _0 \theta _0^2 \eta _{\ast } \left\{ 
\eta _{\ast }^2 \kappa \left( 1 - {1 \over{\eta ^3}} 
\right) \cos \chi + \right. \nonumber \\
& & \left. {3\over 4} \left[ \theta (\eta ) H(\eta) - \theta _0 H(1) \right]
\xi \sin \chi \cos \phi \right\} (1-\xi^2),
\end{eqnarray} 
and the accelerating component of the electric field
\be
E_{\parallel } = - E _0 \theta _0^2 \eta _{\ast }^2 \left[ 
{3\over 2} \eta _{\ast }^2 {{\kappa }
\over {\eta ^4}} \cos \chi +
{3\over 8} \theta (\eta ) H(\eta) \delta (\eta ) \xi 
\sin \chi \cos \phi \right] (1-\xi^2), 
\ee      
where $\Phi _0 \equiv B_0 (\Omega R/c) R f(1)/f(\eta _{\ast })$, 
and $E_0 \equiv \Phi _0/R$.
\noindent
The quantities $\epsilon$ and $\kappa $ in all   
expressions derived in this Section for the rescaled radial 
coordinate now read 
\be
\epsilon \equiv {2GM\over \RE\, c^2},~~~~
\kappa \equiv {\epsilon I\over MR^2}.
\ee
Finally, the functions 
\begin{eqnarray}
&& H(\eta) = {\epsilon\over \eta} - {\kappa\over \eta^3} + 
\left( 1-{3\epsilon\over 2\eta}+{\kappa\over 2\eta^3}\right)
\left[ f(\eta)\left(1-{\epsilon\over \eta}\right) \right] ^{-1}, \\
&& \delta(\eta) = {\partial \over {\partial \eta}} 
\ln{[H(\eta)\theta(\eta)]} 
\end{eqnarray}
are both evaluated at $\eta = r/R_{\rm E}$.

\subsubsection{Screened $E_{\parallel}$ with upper PFF} \label{sec:EPFF}

We now include the effect of the electric field screening at the upper PFF, 
at $\eta =\eta _{\rm c} = (R + h_{\rm c})/R_{\rm E}$, by obtaining the 
solution to Poisson's equation with the upper boundary condition,
$E_{\parallel}(\eta = \eta_c) = 0$. 
In the limit where $z_c \ll 1$ and $z \leq z_c$ the accelerating 
component of the electric field reads
\begin{eqnarray} \label{Ell_1}
E_{\parallel} & \simeq & -E_0\, \theta_0^3\,(1-\epsilon)^{1/2}\,
\eta _{\ast }^2 \left\{ \eta _{\ast }^2 
\left[\sum_{i=1}^{\infty}\,A_i\,J_0(k_i\xi)\right] \cos\chi +  
\right. \nonumber \\
& & \left. \left[ \sum_{i=1}^{\infty}\,B_i\,J_1(\tilde k_i\xi)\right] 
\sin\chi\cos\phi\right\},
\end{eqnarray}
where,
\be \label{A_i,B_i}
A_i = {3\over 2}\,\kappa\,\left[{8\over k_i^4\,J_1(k_i)}\right]\,
{\cal F}_i(z,\gamma_i), ~~~~
~~~~ B_i = {3\over 8}\,\theta_0\,H(1)\delta(1)\,\left[{16\over \tilde 
k_i^4\,J_2(\tilde k_i)}\right]\,{\cal F}_i(z,\tilde \gamma_i),
\ee
\be \label{F(z,gamma)}
{\cal F}(z,\gamma) = -\,[a_1(\gamma\eta-1)e^{\gamma z} + 
a_2(\gamma\eta+1)e^{-\gamma z} + a_1(1-\gamma) -
a_2(1+\gamma)]/(a_1+a_2),
\ee
\be
a_1 = (\gamma\eta_c+1)e^{-\gamma z_c} - \gamma - 1,~~~~~~~~~~
a_2 = \gamma - 1 - (\gamma\eta_c-1)e^{\gamma z_c}, 
\ee
and
\be
\gamma_i \approx {k_i\over \theta_0(1-\epsilon)^{1/2}},~~~~~{\rm and}~~~~~ 
\tilde \gamma_i \approx {\tilde k_i\over \theta_0(1-\epsilon)^{1/2}}, 
\ee
where $k_i$ and $\tilde k_i$ are the positive roots of the Bessel
functions $J_0$ and $J_1$, respectively. In expressions (20), (21) 
$\gamma = \gamma _i$ or ${\tilde \gamma }_i$ should be used. 
In the limit, where $z \equiv \eta - 1 \gg \theta_0 $ and 
$z_c \equiv \eta _c - 1 \gg \theta_0$ (cf. Appendix A, equation [A4]),
\begin{eqnarray} \label{Ell_2}
E_{\parallel} & \simeq & - E_0 \theta _0^2 \eta _{\ast }^2 
\left\{ {3\over 2}\,{\kappa \over {\eta^4}}\,
\eta _{\ast }^2
\left[ (1-\xi^2) - \left( {\eta\over \eta_c}\right)^3
\sum_{i=1}^{\infty}\,{8 J_0(k_i\xi)\over k_i^3\,J_1(k_i)}\,
e^{-\gamma_i(\eta_c)(\eta_c-\eta)}\right] \cos\chi \right. \nonumber \\
&& + {3\over 8}\,\delta(\eta) H(\eta)\theta(\eta)\,\left[ \xi(1-\xi^2)-
{\eta_c \over \eta } {{\theta(\eta_c)\delta(\eta_c)H(\eta_c)} \over 
{\theta(\eta)\delta(\eta)H(\eta)}} \right. \nonumber \\
&& \left. \left. \sum_{i=1}^{\infty}\,{16 J_1(\tilde k_i\xi)\over 
\tilde k_i^3\,J_2(\tilde k_i)}\,e^{-\tilde\gamma_i(\eta_c)
(\eta_c-\eta)}\right] \sin\chi\cos\phi\right\} 
\end{eqnarray}
where now,
\be
\gamma_i = {{k_i}\over {\theta(\eta_c)\eta_c (1-\epsilon/\eta
_{\rm c})^{1/2}}},~~~~~{\rm and}~~~~~
\tilde \gamma_i = {{\tilde k_i}\over {\theta(\eta_c)\eta_c
(1-\epsilon/\eta_{\rm c})^{1/2}}}
\ee
Simple analytic expressions can be derived in several limiting cases
some of which are summarized in Appendix A (for the radial coordinate
scaled by the true stellar radius).  The solutions (\ref{Ell_1}) and
(\ref{Ell_2}) thus
incorporate both the effect of the rescaled lower boundary and
screening of the electric field at the upper boundary, and we will
use them in our numerical calculations described later on in
Section (\ref{sec:numres}).

\subsection{Radiation Production and Losses} \label{sec:rloss}

In this section, we describe our treatment of the radiation processes
that affect the polar cap particle acceleration.  This includes energy
losses due to CR and ICS, and pair production by the photons from these 
processes.  Although the particles are radiating a full spectrum of photons, 
we are concerned here only with the pair producing photons.  In the 
interest of making our numerical code as efficient as possible (but with
some loss of accuracy), we do not model the entire radiation spectrum of
each process, but compute pair production attenuation lengths for a
single representative photon energy at each step along the particle path. 

\subsubsection{Curvature radiation} \label{sec:CR}

The CR loss rate for particles of charge $e$ moving along a
magnetic field with radius of curvature $\rho_c$ is
\be \label{gdot_CR}
-\dot\gamma_{_{\rm CR}} = {3 e^2\over 2mc^2}{c\over \rho_c^2}\gamma^4.
\ee
The radius of curvature of the magnetic field in a Schwarzschild metric is
\be \label{rhoc}
\rho_c \simeq {4\over 3}\left({c\over \Omega r}\right)^{1/2}r~G(x), 
~~~~~{\rm and}~~~~~G(x) = {(1-x)^{3/2}\,f^{5/2}(x)\over 9-2f(x)(4-3x)}
\ee
in the small angle limit, where $x = r_g/r$, $r_g = 2GM/c^2$ is the 
gravitational radius, and the function $f(x)$ has been defined in 
equation (\ref{f}). Formula (\ref{rhoc}) gives a slightly larger 
radius of curvature than the flat-space formula does (by a factor of G, 
about $25-30\%$, for the NS parameters we assume in this paper). 
This can be easily understood, since the strong gravity of the NS tends 
to increase the flaring of the polar field lines, and therefore the 
last open field lines (i.e. those reaching the light cylinder at their 
``turning-points'') should emanate from slightly smaller magnetic
colatitudes. In other words, the polar field lines get slightly
straighten out, and the effective PC radius slightly 
decreases (by the same $\sim 25-30\%$, see also Gonthier \& Harding 1994).

The spectrum of CR photons from a particle with energy $\gamma
$ is a power law at low frequencies, with an exponential decline at 
high frequencies:
\be \label{CRspec}
n_{_{\rm CR}}(\erg) \propto \left\{
\begin{array} {ll} 
\erg^{1/3},  &\erg \ll \erg_{\rm cr} \\
\exp{(-\erg/\erg_{\rm cr})}, &\erg > \erg_{\rm cr} 
\end{array} \right.
\ee
where $\erg_{\rm cr} = (3/2)(\hbar/mc)\gamma^3/\rho_c$ is the critical
frequency. Since the one-photon pair attenuation coefficient 
(equation [\ref{eq:ppasymp}]) increases sharply with energy as 
$\exp{(-8/3\erg B'\sin\theta_{\rm kB})}$, the photons from the 
exponential tail of the curvature spectrum will produce the bulk of
the pairs. Here, and in subsequent Sections, the local dipole magnetic field
strength is
\be \label{B'}
B' = \left({B_0\over B_{\rm cr}}\right)\,\left({R\over R_E}\right)^3
\left[{f(1)\over f(\eta_*)}\right].
\ee
where $B_{\rm cr} = 4.413 \times 10^{13}$ G is the critical field
strength.
The overlap of the curvature spectrum and the pair attenuation
coefficient will fall in a narrow frequency band, which we can approximate in a
steepest descents analysis as a Gaussian of width, 
$\Delta \erg _p = (3B' \erg^3_p \sin \theta _{\rm kB})^{1/2}$, with a mean energy
$\erg_p = (8\erg_{\rm cr}/3B'\sin\theta_{\rm kB})^{1/2}$.  Photons will
pair produce roughly when $\erg_p B'\sin\theta_{\rm kB} \sim 0.2$ for
$B' < 0.1$ and as soon as threshold $\erg_p\sin\theta_{\rm kB} 
= 2$, is reached, for $B' > 0.1$ (Daugherty \& Harding 1983; hereafter
DH83).  
We therefore have
\be
\begin{array}{lll}
\erg_p = 13.3\,\erg_{\rm cr}, &\Delta\erg_p = 0.273\erg_p, &B' < 0.1 \\
\erg_p = 4\,\erg_{\rm cr}/\,3B', 
&\Delta\erg_p = 0.87B'^{1/2}\,\erg_p, &B' > 0.1 
\end{array}
\ee
At each step along the particle acceleration path, pair attenuation 
lengths are 
computed for up to five CR photons with energies evenly distributed 
between $\erg_p$ and $\erg_p + 4\Delta\erg_p$.  The first finite value
(if any) of the pair attenuation length is taken as $S_p$ for that step.

\subsubsection{Inverse Compton scattering} \label{sec:IC}

As the particles are accelerated, they may scatter soft photons from the hot
NS surface.  Due to the strong magnetic field, the ICS 
cross section is resonant at frequencies where electrons may be
excited to higher Landau states and is strongly suppressed (for
extraordinary mode photons) below the fundamental. Although the full
QED scattering cross section for this process has been calculated
(e.g. Daugherty \& Harding 1986, Bussard et al. 1986), it is quite
unwieldy for numerical calculations.  However, the magnetic scattering
cross section in the Thomson limit (Canuto et al. 1971) is much
simpler and has been used in place of the full QED cros section in 
almost all astrophysical applications. It includes only the resonance 
at the cyclotron
fundamental (i.e. scattering in the ground state with no excitation)
and approaches the Thomson cross section at frequencies above the cyclotron
energy.  

In this paper, we follow the treatment of S95, who derived simple
expressions for electron energy loss rates for magnetic Compton
scattering in the Thomson limit, expanding the treatment of D90 to 
include the Klein-Nishina suppression above the
resonance. Dermer's approach was to divide the Compton scattering loss 
rates into three components, each due to the different behavior of the 
cross section in the regions below the resonance (which he called
``angular" scattering due to the strong angle dependence), in the
resonance (``resonant" scattering) and above the resonance
(``non-resonant" scattering).  He assumed that the resonant part of the 
cross section was a
$\delta$-function at the cyclotron energy, and that the soft photon
source was a blackbody of temperature $T_b$ radiated by a hot PC of 
radius $R_T$. He derived the scattering loss rates for
electrons moving away from the NS surface, so that the soft 
photon density decreases with height, due to a dilution factor.
We have expanded this treatment by computing the associated loss rates
for particles (in our case, positrons, although the sign of charge
makes no difference for Compton scattering) moving toward the
NS surface.  In this case, the particles approach the soft photons 
``head-on", scattering them to higher energies than the downward
moving particles.

The Compton scattering energy loss rate for a particle of energy 
$\gamma$ moving through a photon field, $n_{\rm ph}(\erg)$, that
is uniform between angles $\mu_- < \mu < \mu_+$,
from equation (12) of D90, is
\be \label{gdot}
-\dot\gamma_{_{\rm IC}} = {c\over \mu_+ - \mu_-}\,
\int_0^{[\gamma (1-\beta\mu_-)]^{-1}}
d\erg\, n_{\rm ph}(\erg)\,\int_{\mu_-}^{\mu_+}\,d\mu\,(1-\beta\mu)\,
\int_{-1}^1\,d\mu'_s\,\int_0^{\erg'_{s, max}}\,d\erg'_s
\,\left({d\sigma'\over d\mu'_s d\erg'_s}\right)\,(\erg_s - \erg),
\ee
where $\erg'_{s, max} = \gamma\erg(1-\beta\mu)$ is the maximum 
scattered energy in the particle rest frame.  Here, primes are used to 
denote quantities in the particle rest frame and the subscript $s$
denotes scattered quantities.  Thus, $\sigma'$ is the rest frame
scattering cross section.  Based on these expressions, the loss rates
for the three parts of the cross section, $\dot\gamma_{_{\rm IC}} = 
\dot\gamma_{\rm ang}+\dot\gamma_{\rm res}+\dot\gamma_{_{\rm KN}}$,
used in equation (\ref{gdot}) can
be written
\be \label{gdot_ang}
-\dot\gamma_{\rm ang} = 46.1\,T_6^4\,(1-\mu_c)\,f_{\rm
ang} ~~~{\rm s^{-1}},
\ee
\be \label{gdot_res}
-\dot\gamma_{\rm res} = 4.9 \times 10^{11}\,
\left[{T_6\,B_{12}^2\over \beta\gamma} \right]\,f_{\rm res}~~~{\rm s^{-1}} ,
\ee
\be \label{gdot_KN}
-\dot\gamma_{_{\rm KN}} = 3.7 \times 10^{11}\,
\left({T_6\over \beta\gamma^2}
\right)\int_{\erg'_-}^{\erg'_+}\,d\erg'\,[\erg_0 +
\erg_0/\erg' - \gamma]\,f_{\rm KN} ~~~ {\rm s^{-1}},
\ee
where $T_6 \equiv T_b/10^6$ K is the NS surface temperature, $B_{12}$
is the local dipole magnetic field strength in units of $10^{12}$ G. 
equation (\ref{gdot_KN}) assumes a $\delta$-function at
energy $\erg_0 = 2.7\theta_{\rm T}$, where $\theta_{\rm T} = kT/mc^2$,
for the distribution of soft thermal photons, and
\be \label{fang}
f_{\rm ang} = {{\gamma^2-2\over \gamma^2-1} - {(\mu_+^3-\mu_-^3)\over 
3(\mu_+ -\mu_-)} - {(\mu_+ + \mu_-)\over 2\beta\gamma^2} + 
{\ln{[(1-\beta\mu_-)/(1-\beta\mu_+)]}
\over \beta^3\gamma^4(\mu_+ - \mu_-)}} ,
\ee
\begin{eqnarray} \label{fres}
f_{\rm res} & = &\left( 1-{\eB\over 3}\right) \ln{\left[{1-e^{-w_+}
\over 1-e^{-w_-}}\right]} + \left({2\over 3}-{1\over \eB}\right) {1\over \gamma}
\nonumber \\
&& \left\{[w_+
\ln(1-e^{-w_+})-w_-
\ln(1-e^{-w_-}) - \theta_T[Li_2(e^{-w_+}) - Li_2(e^{-w_-})]\right\},
\end{eqnarray} 
\begin{eqnarray} \label{fKN}
f_{\rm KN}& = &{2(\erg'^2-4\erg'-3)\over \erg(1+2\erg')} +
{2\erg'\over 3}{\left[4\erg'^2+6\erg'+3
\over (1+2\erg')^3\right]} - 
\nonumber \\
&&{2(1+\erg')(\erg'^2-2\erg'-1)\over \erg'(1+2\erg')^2} -
\left[{\erg'^2-2\erg'-3\over \erg'^2}\right]\ln{(1+2\erg')} - 
{2\over \erg'} .
\end{eqnarray} 
In the above expressions, $w_{\pm} = \eB/[\theta_{\rm T}
\gamma (1-\beta\mu_{\pm})]$, $\erg'_{\pm} = \erg_0
\gamma(1-\beta\mu_{\pm})$ and $\eB = B'$ is the local cyclotron energy 
in units of $mc^2$. Equations (\ref{gdot_ang}), (\ref{gdot_res}) and (\ref{gdot_KN})
are identical to those given by S95, but the expressions
for $f_{\rm ang}$, $f_{\rm res}$ and $f_{\rm KN}$ have been
generalized to allow for scattering by downward moving
particles.  Specifically, the corresponding equations given by S95 have
specialized to the case of upward moving particles ($\mu_- = \mu_c,
\mu_+ = 1$, see equation~(\ref{mupm}) below).
equation (\ref{fang}) is the same as D90's equation (16), 
which is general enough to compute the loss rate of either upward or 
downward moving particles. Equation (\ref{fres}) includes, to first order, 
the rest-frame particle recoil $\erg'_s \simeq \erg'[1 - 
\erg'(\mu'_s-\mu')^2/2]$ and is a major improvement over previous 
expressions for resonant Compton scattering that assume no recoil
($\erg'_s = \erg'$).  The first term in equation (\ref{fres}) gives
D90's equation (54), and the remaining terms result from recoil, 
where $Li_2$ is the Dilogarithm function.  The treatment of recoil in 
resonant scattering is necessary in this calculation to compute
accurate energies for photons scattered by the electrons and positrons.  

Given the generality of the above expressions, the only difference 
in computing scattering loss rates for upward-moving electrons and 
downward-moving positrons lies in the integration limits over polar
angle $\mu$ in the laboratory frame.  For the case of semi-isotropic
blackbody radiation at height $h = s + h_0$ above a hot PC of 
radius $R_{\rm T}$, this translates into the following values for
\begin{eqnarray}  \label{mupm}
\mbox{electrons:}&&\mu_- = \mu_c \nonumber \\ 
&&\mu_+ = 1 \\
\mbox{positrons:}&&\mu_- = -1 \nonumber \\
&&\mu_+ = -\mu_c 
\end{eqnarray}
where
\be
\mu_c = \cos\theta_c = {h\over \sqrt{h^2 + R_{\rm T}^2}},
\ee
(when $R_{\rm T} \ll R$) gives the dilution factor that is present in the soft 
photon density,
\be \label{n_ph}
n_{\rm ph}(\erg) = n_{\rm bb}(\erg)\left({1-\mu_c\over 2} \right)
\ee
where
\be
n_{\rm bb}(\erg) = {8\pi \over \lambda_C^3}{\erg^2\over 
  [\exp{(-\erg/\theta_{\rm T})}-1]}
\ee
is a blackbody spectrum of temperature $\theta_{\rm T}$ and 
$\lambda$ is the electron Compton wavelength.

To determine the energy of the scattered photons that may produce pairs, the 
scattering kinematics of the electron and positron must be treated accurately.
For non-resonant scattering in the Thomson limit, i.e. 
when $4\gamma\erg_0 < 1$, we can assume there is no recoil in
the particle rest frame, and the maximum scattered energies in the 
laboratory frame are
\be
\begin{array} {ll}
\erg_s = 2\gamma^2(1-\mu_c)\erg_0   &\mbox{electrons},\\
\erg_s = 4\gamma^2\erg_0   &\mbox{positrons}.
\end{array}
\ee
Since the electrons and the soft photons are both moving away from the 
NS surface, 
$\erg_s$ depends on $\mu_c$, which decreases with height.  However, the
positrons scatter the soft photons ``head-on" with a maximum angle of $\pi$,
and $\erg_s$ is independent of height.
When $4\gamma\erg_0 > 1$, or for resonant scattering, we must include 
recoil in the rest frame
to compute the scattered energies.  The average scattered energy in
the laboratory frame is computed from the scattering rates, which for 
$\dot\gamma_{\rm res}$ (equation [\ref{gdot_res}]) and $\dot\gamma_{_{\rm
KN}}$ (equation [\ref{gdot_KN}]) explicitly include recoil, and the loss
rates, so that
\be
\langle \erg_s \rangle \approx {\dot\gamma_{_{\rm IC}}\over R_{\rm IC}}
\ee
for each component, where
\be \label{RIC}
R_{\rm IC} = {c\over \mu_+ - \mu_-}\,\int_0^{[\gamma (1-\beta\mu_-)]^{-1}}
d\erg\, n_{\rm ph}(\erg)\,\int_{\mu_-}^{\mu_+}\,d\mu\,(1-\beta\mu)\,
\int_{-1}^1\,d\mu'_s\,\int_0^{\erg'_{s, max}}\,d\erg'_s
\,\left({d\sigma'\over d\mu'_s d\erg'_s}\right).
\ee
$\langle \erg_s \rangle$ is thus the weighted average of 
$(\erg_s - \erg)$, where $(\erg_s \gg \erg)$ for scattering
by relativistic particles.  Since photons scattered by the ``angular" part of
the cross section below the resonance do not have energies above pair 
threshold, for our calculation we need only $R_{\rm res}$ and $R_{\rm KN}$. 
Using equation (\ref{RIC}) and following Dermer's (D90) method of computing 
$\dot\gamma_{\rm res}$, we have
\be
R_{\rm res} = 2.18 \times 10^{13}
\left({T_6\,B_{12}\over \beta\gamma^2} \right)\,
\ln{\left({1-e^{-w_+}\over 1-e^{-w_-}}\right)}~~~{\rm s^{-1}} .
\ee
For $\gamma \gg 1$, the resonant scattered photon energy, $\erg_s^{\rm res}
\simeq \gamma B'$ is independent of incident angle, and is thus the same
for upward and downward moving particles.  However, the energy of the
incident photon in the particle rest frame, $\erg'_{\pm} = \erg_0
\gamma(1-\beta\mu_{\pm})$ will not be the same, so the upward and downward
moving particles will not necessarily both undergo resonant scattering.
Using equation (\ref{RIC}) and following
Sturner's (S95) method of computing $\dot\gamma_{\rm KN}$, we have
\be
R_{\rm KN} = 3.7 \times 10^{11} \left({T_6\over 
\beta\gamma^2} \right)\int_{\erg'_-}^{\erg'_+}\,d\erg'\,r_{_{\rm KN}} 
~~~ {\rm s^{-1}}, 
\ee
where
\be
r_{_{\rm KN}} = {4\over \erg'} + {2\erg'(1+\erg')
\over (1+2\erg')^2} + {(\erg'^2-2\erg'-2)
\over \erg'^2}\ln{(1+2\erg')}.
\ee
Our numerical calculations show that $\langle \erg_s^{\rm KN} \rangle$ 
approaches $\gamma$ for $4\gamma\erg_0 \gg 1$, as expected.
Thus, $\langle \erg_s^{\rm KN} \rangle > \langle \erg_s^{\rm res} \rangle$ 
for $B' < 1$.

\subsection{Pair Production} \label{sec:pair}

The method we use to compute the electron-positron pair attenuation
length of the photons, $S_p(\erg)$, has been described in detail
in Harding et al. (1997). Using equation (\ref{tau}), $S_p(\erg)$ is 
computed by integrating the pair production attenuation coefficient of 
the photon along its path through the dipole field. The photon is
assumed to pair produce at the point where $\tau(\theta, \erg) = 1$, 
and $S_p(\erg)$ is then set to that path length.  The two main
inputs needed are the energy of the photon and its angle to
the magnetic field, $\theta_{\rm kB}$. The energies of the radiated
photons were given in Sections \ref{sec:CR} and \ref{sec:IC}.  For 
the angle of the radiated photons at the emission point, 
we assume that $\theta_{\rm kB}
= 2/\gamma$ for Compton scattered photons (D90) and 
$\theta_{\rm kB} = 0$ for CR photons.  Due to the 
curvature of the field lines, $\theta_{\rm kB}$ will grow as the
photon propagates, roughly as $\sin\theta_{\rm kB}
\sim s/\rho_c$.  To evaluate the pair production attenuation coefficient
at each point along the path of the photon, we Lorentz-transform the 
photon energy and local magnetic field to the frame in which the 
photon propagates perpendicular to the local field.  This is the 
center-of momentum frame for the created pair, where the attenuation
has its simplest form and the photon energy is $\erg_{_{\rm CM}} = 
\erg\sin\theta_{\rm kB}$.  The one-photon pair attenuation coefficient is
considered in two regimes.  For $B < 0.1\,B_{\rm cr}$, photon pair
produce far above threshold, where the asymptotic expression in the
limit of large numbers of kinematically available pair
Landau states (Tsai \& Erber 1974, DH83) can be used:
\be   \label{eq:ppasymp}
   T^{\rm pp}_{\parallel,\perp} = {1\over 2}{\alpha\over \lambar} B'
   \Lambda_{\parallel,\perp}(\chi),
 \ee

\be 
   \Lambda_{\parallel, \perp}(\chi) \approx \left\{
     \begin{array}{lr} 
     (0.31, 0.15)\, \exp \mbox{\Large $(-{4\over 3\chi})$} & \chi \ll 1 
     \\ \\
     (0.72, 0.48) \, \chi^{-1/3} & \chi \gg 1
     \end{array} \right.          \label{eq:ppratlim}
\ee
where $\chi \equiv f\,\erg_{_{\rm CM}}/2B'$, $\alpha$ is the 
fine-structure constant, $B'$ is the dipole
field strength at point $s$ along the photon path (see equation [\ref{B'}]).  
When $B > 0.1\,B_{\rm cr}$, pair production will occur near threshold,
where the above expression is not accurate.  We thus include the
factor $f = 1+0.42\erg_{_{\rm CM}}^{-2.7}$ in $\chi$, introduced
by DH83, as an approximation to the near-threshold attenuation 
coefficient.
In this paper we compute the attenuation length averaged over photon 
polarization.

\subsection{Backflowing Positrons, Double Pair-fronts and
 Polar Cap Heating} \label{sec:pfraction}

Soon after the concept of electron-positron pair production had been 
introduced into pulsar models (Sturrock 1971), it became clear 
that the precipitation of positrons (or, in some models, electrons) 
onto the stellar surface may have some important effects. For example, 
Sturrock has noticed that (see p. 531) ``if the positrons were all 
returned to the
surface, the resulting space charge would reverse the sign of the 
electric field at the surface, cutting off the flux of primary
electrons''. Then he added that ``one possibility is that the
configuration so adjusts itself that the primary flux is never quite
cut off, so that the flow is steady, and another possibility is that
the flow is oscillatory''. Thus, he drew attention to the existence of 
an intrinsic feedback between the pair formation and the
electric field accelerating primary particles. Later on, several 
authors (see e.g. RS75, Jones 1978, AS79, and Arons 1981) have
discussed similar ideas
focusing on the possibility of PC heating by the flow of
energetic positrons (electrons). DH96 discussed the effect of downward
moving positrons on the particle acceleration and the possibility of a
lower PFF.  The most detailed study has been
performed by AS79 who calculated the fraction
of returning positrons, structure of the PFF with positron trapping 
near but below the lower boundary of the PFF, and briefly discussed 
the positron-initiated cascade very near the stellar surface. They 
estimated the fraction of returning positrons as $\sim \theta_0^2$  
($\ll 1$, where $\theta _0$ is the half-opening angle of the polar flux tube
at the stellar surface) of the Goldreich-Julian charge density. The 
main argument underlying this estimate is that the contribution of
the second term in the square bracket in equation (4) increases with 
altitude (due to the flaring of the magnetic field lines emanating
from the PC), and that (see AS79, p. 867) 
``the additional negative charge 
density needed to achieve $E_{\parallel }=0$, over and above that of 
the electron beam, is therefore a fraction $\sim \theta _0^2$ of 
the negative charge density already present in the electron
stream''. And, ``as the pairs begin to form near but below some 
height (at which the electric field shuts off), the residual
$E_{\parallel }$ of the diode causes the secondary electron-positron
plasma to polarize and shield these ``external'' charges. The
formation of this polarization charge requires a dynamical response
which, under most conditions, leads to the formation of the downward 
directed positron stream whose flux at the stellar surface is 
$F_+ \ll F_-$''. However, the main problem with this reasoning is that 
if we assume the zero-electric field boundary condition at the stellar 
surface, then ``the additional negative charge density'' is nothing
else but the imbalance between the real and Goldreich-Julian charge 
densities that produces the electric field above the stellar surface
in the first place. This imbalance (or charge deficit) increases with 
altitude and reaches some maximum value. Thus, if the backflowing 
positrons shield these ``external'' charges at some height, then 
they will also screen the electric field all the way down to the
stellar surface, and we 
should unavoidably come up with the situation envisaged by Sturrock, 
where the returning positrons cut off the flux of primary electrons, 
thus resulting in the oscillatory regime. 

In this section, we explore the possibility of occurrence of a steady regime of
positron backflow that is not capable of disrupting the current of
primary electrons. We also discuss the efficiency of PC  
heating produced by the precipitating relativistic positrons and 
high-energy quanta.  We also discuss the accounting of the feedback
between the returning positron flux and electric potential
distribution in the acceleration region in the self-limiting 
regime of particle flow. The primary (electron) traversing the region 
of the electric potential
drop emits a $\gamma -$ray photon that produces an electron-positron 
pair at the altitude $h_{\rm pair}$ above the stellar surface. The
electron-positron pairs keep moving forward, with some positrons of these
pairs decelerating and eventually reversing the direction of their motion 
at some altitude $h_{\rm c}$. It is very likely that the same effect 
tends to occur with the positrons from both additional photons emitted 
by the same electron or from a second generation of pairs. An accurate 
determination of the
fraction of backflowing positrons would require a self-consistent
calculation of the screening of $E_{\parallel }$ by the cascade pairs
and the modification of the pair energy and spatial distribution 
by the screened 
$E_{\parallel }$. Short of carrying out such a detailed calculation
here, we shall derive an upper limit on the fraction of returning
positrons. However, since the cascades develop very 
fast and the number density of particles increases in an avalanche, 
the altitudes at which the positrons from the higher-order cascades 
get decelerated should be very close to but slightly above $h_{\rm
c}$. The deceleration of positrons results in a charge separation in 
the initially quasi-neutral electron-positron beam, thus effectively shielding
the electric field where the bulk of the pair front is produced. We
can justifiably assume that $E_{\parallel }=0$ at 
$h \approx h_{\rm c}$. The backflowing positrons slightly suppress the 
voltage all the way down to the bottom of the polar magnetic flux tube, 
so that the electric field $E_{\parallel }$ (and also potential $\Phi
$) vanishes at a height greater than the start-off height of the primary electrons. This occurs because the flux of 
backflowing positrons is equivalent to a corresponding enhancement 
of the total electron current and therefore of the maximum number of 
electrons per second to be ejected into the acceleration region. 
Thus, in compliance with the zero-electric field boundary condition, the 
ejection radius should fix itself at the larger value corresponding 
to a higher effective ``rate of supply of Goldreich-Julian charge'' 
(see also expressions [4] and [\ref{IetaE}] proving this statement). 

Let us introduce the Goldreich-Julian current (associated with the 
corresponding charge density) or the rate of ejection of charges 
per element of the solid angle into the region of open magnetic field lines 
\be
{{d{\cal I}_{\rm GJ}}\over {d \Omega _{\xi }}} = \alpha c 
| \rho _{\rm GJ} | (r\theta )^2,
\ee
where $(r\theta )^2 \equiv (\Omega R/c) 
R^2 \eta ^3 \xi ^2 /f(\eta )$, and 
$d \Omega _{\xi } = \xi d \xi d \phi $ is an element of the
solid angle in the PC region. In the rest of this Section we
present our estimate of the maximum power carried by the backflowing 
positrons. Using the general-relativistic expression
for $\rho _{\rm GJ}$ (equation [\ref{rhoGJ}]) we can write
\be  \label{IetaE}
{{d{\cal I}_{\rm GJ}(\eta _{_{\rm E}})}\over {d \Omega _{\xi }}} = 
{{{\cal I}_0}\over {2\pi }} \left[ \left( 1 - {{\kappa}\over {\eta
_{_{\rm E}}^3}} \right)
\cos \chi + {3\over 2} H(\eta _{\rm E}) 
\theta (\eta _{_{\rm E}}) \xi \sin \chi \cos \phi \right],
\ee     
where ${\cal I}_0 \equiv (\Omega R/c)^2 B_0 R c/f(1)$ and $\eta _{_{\rm E}} \equiv R_{\rm E}/R \equiv (R+h_0)/R=1+h_0/R >
\eta$.

The zero-electric field boundary condition at $\eta = 1$ requires 
the magnitude of the primary electron current flowing from the PC 
surface to be equal to $(d {\cal I}_{\rm GJ}/d \Omega _{\xi })_
{|_{\eta =\eta_{_{\rm E}}}}$. 
This expression implies (compare with expression [4]) that an enhanced 
electron current is ejected 
from the effective radius in the regime of self-limitation or,
equivalently, allows for the additional current of positrons flowing  
from the upper PFF downward. 
It is important that in the steady-state regime 
the total current of actual charges
(electrons and positrons) remains constant along the field lines, and
is fixed by the Goldreich-Julian current at the zero-electric 
field boundary (i.e., in our case, at the effective radius $R_{\rm E}$). 
The electric field above $R_{\rm E}$ (in the regime of self-limitation) 
is produced by the imbalance between the actual current and the local
value of the Goldreich-Julian current. At distances greater than the
effective radius $R_{\rm E}$ for the Goldreich-Julian current we can 
write
\be  \label{Ieta}
{{d{\cal I}_{\rm GJ}(\eta > \eta _{\rm E}) 
}\over {d \Omega _{\xi }}} \approx  {{{\cal I}_0}\over {2\pi }} 
\left[ \cos \chi + {3\over 2} H(\eta ) 
\theta (\eta ) \xi \sin \chi \cos \phi \right].
\ee

Let us consider the situation when the fraction of positrons
returning from the upper PFF is not only sufficient to produce a
lower PFF, but is also big enough to affect the electric field 
accelerating the primary electrons. For example, the backflowing 
positrons tend to
reduce the accelerating electric field above the effective surface
of radius $R_{\rm E}$, simply because the positron current is
equivalent to the enhanced electron current which unavoidably requires 
the electrons to be ejected from the greater heights above the
effective surface. Thus, the returning positrons may eventually
screen the accelerating electric field and trigger an oscillating
regime with alternating phases of developing and collapsing of double
pair fronts. As has been mentioned above, the oscillatory regime 
akin to that we discuss here was anticipated in his classical 
paper by Sturrock (1971), even though he did not discuss the lower
PFF. The 
occurrence of the oscillatory regime depends on the fraction of
backflowing positrons, which is determined by the particle  
kinematics within the upper PFF and the penetration depth of the 
electrostatic field into the electron-positron plasma cloud. A 
quantitative description of this process would be possible through 
a detailed analysis of the electron-positron cascades and their
feedback on the electrodynamics of the pair formation region. 
We can now estimate the maximum positron current needed to screen the 
electric field accelerating the primary electrons. For this purpose 
we assume that the 
bulk of the lower pair front sets up at the effective radius 
$R_{\rm E}$, and that the upper pair front establishes at the radial
distance $r > R_{\rm E}$. Then, for the maximum positron current we
can write, using equations (\ref{IetaE}) and (\ref{Ieta}),
\begin{eqnarray}
\left( {{d{\cal I}_{\rm e^+}}\over {d \Omega _{\xi }}} 
\right)_{\rm max} & \approx & {d\over {d \Omega _{\xi }}}
\left[ {\cal I}_{\rm GJ} (\eta > \eta _{\rm E})-
{\cal I}_{\rm GJ} (\eta _{\rm E})
\right] = \nonumber \\
& & {{{\cal I}_0}\over {2\pi }} 
\left[ \kappa \eta _{\ast }^3 \cos \chi + 
{3\over 2} H(\eta ) \theta (\eta ) \xi \sin \chi \cos \phi \right],
\end{eqnarray}
where $\eta _{\ast }\equiv R/R_{\rm E}$.
The maximum value of the electric potential can be  estimated as 
(cf. equation [13])
\be
\Phi _{\rm max} \approx {{{\cal I}_0} \over {2c}} \eta _{\ast } 
\left[ \kappa \eta _{\ast }^2 \cos \chi + {3\over 4} H(\eta ) 
\theta (\eta ) \xi \sin \chi \cos \phi \right] (1-\xi ^2).
\ee
Now we can derive the maximum total power put into the 
backflowing positrons (cf. MH97, equations 
[76]-[78]):
\be
\left\{ L_{\rm e^+} \right\} _{\rm max } 
\approx \int_{\Omega _{\xi }}  \Phi _{\rm max} \left\{ 
d{\cal I}_{\rm e^+} \right\} _{\rm max} \approx 
\left\{ \lambda _{+}\right\} _{\rm max}~{{{\cal I}_0^2} \over {6 c}} \equiv 
\left\{ \lambda _{+}\right\} _{\rm max}~L_{\rm sd},
\ee
where 
\be    \label{lam+}   
\left\{ \lambda _{+} \right\} _{\rm max} = {3\over 4} \eta _{\ast } \left( 
\kappa ^2 \eta _{\ast }^5 \cos ^2 \chi + {3\over 16} H^2 \theta ^2 \sin ^2
\chi \right),
\ee
and
\be
L_{\rm sd} \equiv {{\Omega ^4 B_0^2 R^6}\over {6 c^3 f^2(1)}} 
\ee
is the spin-down luminosity of a pulsar. A similar estimate for the
maximum efficiency of acceleration of primary electrons yields 
\be      \label{lam-}
\left\{ \lambda _{-} \right\} _{\rm max} = {3\over 4} \left[ 
\kappa \eta _{\ast }^3 \left( 1 - \kappa \eta _{\ast }^3 \right) 
\cos ^2 \chi + {3\over 16} H^2 \theta ^2 \sin ^2
\chi \right],
\ee
which amounts to 10 $\%$ (for a small obliquity and $\eta _{\ast
}\approx 1$). 

For relatively small obliquities (or small magnetic polar angles and 
non-orthogonal rotator) and typical pulsar spin periods of 
0.1 - 1 s the first term in equation (\ref{lam+}) dominates, and we get 
\be  \label{lam+max}
\left\{ \lambda _+ \right\} _{\rm max } \approx 
{3\over 4} \kappa ^2 \eta _{\ast }^6 .
\ee

For a 1.4 solar mass NS and for a broad range of realistic
equations of state of dense matter (see e.g. Lorenz, Ravenhall, \& 
Pethick 1993 and Ravenhall \& Pethick 1994 for the calculations of the
NS moment of inertia for various equations of state) 
$I/(MR^2) \approx (0.2-0.25) (1-r_{\rm g}/R)^{-1}$. Thus, for
the NS of 1.4 solar mass and 8-10 km radius (which is
consistent with the most realistic stellar models) we can estimate 
$\kappa \equiv (r_{\rm g}/R)(I/MR^2) \approx 0.15-0.27$. Given 
$\eta _{\ast } \simeq 0.5-1$ (see Section \ref{sec:numres}), equation 
(\ref{lam+max}) yields
\be   \label{lam+num}
\left\{ \lambda _+ \right\} _{\rm max }  
\approx \left(3\cdot 10^{-4}-2\cdot 10^{-2} \right) 
\left( {{\kappa }\over 0.15} \right)^2 .
\ee
This estimate combined with the recently observed X-ray luminosities (that
include both pulsed and non-pulsed components and imply isotropic
emission) of pulsars (see Becker \& Tr\"{u}mper 1997) may have rather
interesting implications. If the X-ray fluxes in some of 
these pulsars are dominated by the photons from the heated 
(e.g. by the backflowing positrons) PC, then luminosities higher
than given by equation (\ref{lam+num}) would
indicate that these pulsars operate in the oscillatory regime 
discussed above.  According to the 
estimates by Becker \& Tr\"{u}mper the 
pulsed X-ray luminosities (for the case of isotropic emission), 
$L_{\rm x}^{\rm p}$, for e.g. Crab, Vela, Geminga, and PSR 0656+14 are, 
respectively, $1.6\cdot 10^{-3}$, $0.7 \cdot 10^{-5}$, 
$1.3\cdot 10^{-4}$, and $3.7\cdot 10^{-3}~L_{\rm sd}$ (where 
$L_{\rm sd} \equiv I \Omega |\dot{\Omega }|$, $\Omega $ and $\dot{\Omega
}$ are the pulsar spin frequency and its time derivative,
respectively).  Although some of these pulsars have X-ray luminosities
that lie above $\left\{ L_{\rm e^+} \right\} _{\rm max }$, the X-ray emission may not be isotropic (see Zavlin et al. 1995).  If we assume that at 
least for Vela, Geminga, and
PSR 0656+14 (see e.g. Harding \& Muslimov 1998 for the modeling of
the soft X-ray and $\gamma-$ray emission for these pulsars) the X-ray 
emission is beamed into a solid angle of $\sim $ 1 steradian, then for
these pulsars we can estimate that ${\rm \{\lambda _x^p\}_{anis}} 
\equiv \{L_{\rm x}^{\rm p}\}_{\rm anis}/L_{\rm sd} \rm \sim 6 
\cdot 10^{-7}$, $10^{-5}$, and $3\cdot 10^{-4}$, respectively. The 
corresponding PC temperature for these pulsars can be estimated as 
\be
T_{\rm pc} \sim (0.6-1) \cdot 10^6 \left( 10^5~\{\rm \lambda
_x^p\}_{\rm anis} \right) ^{1/4} 
\left( {{B_0}\over {4\cdot 10^{12}~{\rm G}}} \right) ^{1/2} 
\left( {R\over {\rm 8~km}} \right) ^{3/4} 
\left( {P\over {\rm 0.1~s}} \right) ^{-3/4} ~{\rm K}.
\ee
Thus, the estimated value of $\rm \{\lambda _x^p\}_{anis} \sim (3\cdot 
10^{-5}-10^{-2}) \left\{ \lambda _+ \right\}_{max}$ may indicate that 
the backflowing
positrons precipitate onto the effective area smaller than that of the
standard PC (e.g. the returning positrons focus toward the 
magnetic axis as discussed below), and/or that the fraction of the
returning positrons is well below the maximum possible one. Then the 
latter would support the quasi-steady state (MH97) rather than the 
oscillatory (Sturrock 1971) regime of pulsar operation.

The energetics of the PC heating is mainly determined by the
energetics of the backflowing (primary) positrons. Their energy is 
eventually redistributed between the high-energy photons they emit 
and electron-positron pairs created by photons in the magnetic
field. Let us consider the last open field line of the dipole magnetic 
field. The photons emitted by backflowing positrons produce pairs on
adjacent innermost field lines, which may reduce the effective area of
the PC subject to the heating. Here we shall present very rough estimate
of the geometrical filling factor for the PC heating qualitatively 
illustrating this effect. The photons emitted by backflowing positron
(moving along the last open field line) at height $S_{\rm c}^+$ above
the effective radius $R_{\rm E}$ produces a pair at the effective 
radius and at the angular distance from the magnetic axis 
\be
r^{\ast } = (R_{\rm E}+S_{\rm c}^+)\tan \theta - S_{\rm c}^+ \tan
\lambda ,
\ee
where $\theta $ is the magnetic colatitude of the point at which the
photon is emitted, and $\lambda $ is the angle between the tangent to
the field line at the point of photon emission and the normal to the 
stellar surface (here we may justifiably neglect the surface
curvature). In a flat space limit 
elementary geometrical consideration yields the relationship
$\lambda = \mu + \theta $, where $\mu $ is the angle between the
tangent to the field line and the radius-vector of the point on the
field line with the magnetic colatitude $\theta $. For a dipole field, 
$\tan \mu \approx {1\over 2} \tan \theta \approx {1\over 2} \theta $,
so that $\tan \lambda \approx {3\over 2} \theta $, and we get 
\be
r^{\ast } = R_{\rm E}\left( 1-{{S_{\rm c}^+}\over {2R_{\rm E}}}
\right) \theta .
\ee
Since $\theta = \theta _{\rm pc} [(R_{\rm E} + S_{\rm c}^+)/R]^{1/2}$, 
where $\theta _{\rm pc}$ is the angular size of a standard PC, we can
write for the effective angular size (at the actual stellar radius) of 
the heated PC 
\be
\theta ^{\ast } = \theta _{\rm pc} \left( 1+{{S_{\rm c}^+}\over {R_{\rm E}}}
\right) ^{1/2} \left( 1-{{S_{\rm c}^+}\over {2R_{\rm E}}} \right) .
\ee
Thus, in the case where $S_{\rm c}^+ \sim R$, and $R_{\rm E}\sim
(1-2)R$, we get 
$$
\theta _{\rm pc}^{\ast } \approx (0.7-0.9) \theta _{\rm pc} . 
$$
The effective area of the heated PC may therefore be up to 50 $\%$
less than the area of a standard PC. The above estimate can be
additionally justified by the fact that $S_{\rm c}^+$ increases toward 
the PC rim, while $R_{\rm E}$ remains practically constant. We thus 
suggest that for some pulsars this effect may be worthy of discussion.

\section{Self-Limited Acceleration Zone}

\subsection{Numerical Results} \label{sec:numres}

Following the procedure outlined in Sections \ref{sec:Ell}-\ref{sec:pair}, 
we have made numerical calculations of the PFFs and the total acceleration 
length, $S_c$, from equation (\ref{Sc}), for both
upward moving electrons and downward moving positrons. 
The parameters of the thermal radiation from the NS surface are
somewhat uncertain.  Although measured temperatures of pulsed X-rays 
lie in the range $5 \times 10^5 - 10^6$ K, the distribution of the 
radiation is not known.  For the calculations of this Section,
we have assumed that the radiation is isotropically emitted from a hot 
PC of radius $R_T = 3 \theta_0 R$ and that $R = 8$ km. 
Other possibilities, and how they might affect these results will be 
discussed in Section \ref{sec:Dis}. An electron is started at
height $h_0$ above the NS surface with $\gamma^- = 1$.  
For a given value of $h_0$ and thus, $R_E = R + h_0$, a ``first guess" value 
of $h_c$, and thus also of $S_c^-$, sets the initial acceleration
length. equation (\ref{dgamma}) is integrated in discrete steps
upward from the starting point, computing $E_{\parallel}$ (from either
equation [\ref{Ell_1}] or [\ref{Ell_2}]), 
$\dot\gamma_{_{\rm IC}}$ and $\dot\gamma_{_{\rm CR}}$ at each step.  
At each step, the pair attenuation lengths, $S_p(\erg)$, of both CR and 
ICS test photons radiated by the particle of energy
$\gamma(s)$ are computed from equation (\ref{tau}).  The pair
attenuation length, and thus the value of $S_c^-$, also computed 
at every step, is initially infinite, because the energy of the photons 
is small, but decreases with distance as the energy of the radiated
photons increases.  Although the photon attenuation length continues to
decrease, the particle acceleration length is increasing and $S_c^-$ has 
a minimum.  This minimum value of $h_c = h_0 + (S_c^-)_{\rm min}$ is 
adopted as the new value of the
electron PFF for the assumed value of $h_0$.  The electron is
accelerated again with the new value of $h_c$, producing a new PFF at the
next value of $h_c = h_0 + (S_c^-)_{\rm min}$.  
The process is repeated, converging to a self-consistent value of $h_c$.  

When $h_0 \ll R$, we find that the value of $S_c^-$ for ICS photons is
smaller than that for CR photons.  PFFs at low altitudes are therefore 
ICS-controlled.
Figure \ref{fig:prim1} shows examples of self-consistent solutions
of the ICS-controlled electron PFF near the NS surface (left panel) and
the corresponding ICS-controlled positron PFF (right panel).  
In the left panel of Fig. \ref{fig:prim2},
an electron starts very near the NS surface (at height $h_0 = 0.01\,R$), 
where the density of soft thermal photons and the loss rate due to ICS is high.  In this case, photons from resonant ICS produce pairs well before CR photons 
and thus the 
ICS photons establish the PFF.  In fact, the loss rate for CR is
orders of magnitude smaller than that for ICS at the PFF.  In the right
panel of Fig. \ref{fig:prim1}, the loss rates and integrated energy of the
positron, accelerated by the same electric field as the electron, is shown as
a function of the acceleration length.  It this case, however, the larger
angles between the thermal X-ray photons and the positrons allow them
to accelerate through the resonant part of the ICS losses at lower
energies, and non-resonant ICS of photons above the cyclotron energy produce 
the PFF.  Since at pulsar field strengths, scattering above the cyclotron
energy occurs in the KN regime, the scattered photon energies are much larger
than those of photons scattered in the resonance.  Therefore, the ICS photons
producing the positron PFF have higher energies than the ICS photons producing
the electron PFF (see Fig \ref{fig:epsmin}) and the positron PFF forms in a shorter
distance.  This is the major cause of the difference between the 
electron and positron ICS-controlled PFFs,
which will be discussed in more detail in connection with Fig. \ref{fig:pffep}.
Figure \ref{fig:prim2} shows an example of the self-consistent solution for 
a CR-controlled electron PFF.  The electron
begins accelerating at a significant fraction of a stellar radius above 
the surface, where the soft photon density has dropped, CR losses dominate and
CR photons establish the PFF.  Because CR is much less efficient than ICS, 
the particles must accelerate to higher energies to produce pairs and the 
total acceleration length $S_c$ is longer.

We first explore the solutions for the electron PFF, as a function of 
pulsar parameters, for acceleration from the surface.  Figures 
\ref{fig:pcacch} and \ref{fig:pffxb} show the height of the electron 
PFF and the maximum acceleration voltage ($\gamma_{\rm max}$) as a function 
of the scaled colatitude $\xi$ for CR-controlled 
PFFs, assuming that ICS is ``turned off". These results can be
directly compared with Fig. 5 of A83, who computed the PFF due only 
to curvature photons.  Figure \ref{fig:pcacch} shows the strong effect 
of obliquity $\chi$ on the shape of the electron PFF and on the acceleration 
voltage, due primarily to the dependence of $E_{\parallel}$ on $\chi$.  
Since the frame-dragging component of $E_{\parallel}$ proportional 
to $\cos\chi$ (cf. equation [\ref{Ell_1}]) is much stronger near the 
surface than the component proportional to $\sin\chi$, it dominates at 
small and intermediate values of $\chi$.  The frame-dragging
electric field becomes comparable to that in a flat space for orthogonal 
rotators ($\chi = 90^0$), where our solutions match very well with
those of A83's Fig. 5.  The height of the 
CR-controlled PFF ($S_c^-$) increases and $\gamma_{\rm max}$ decreases with
increasing $\chi$, as $E_{\parallel}$ decreases.  
The height of the PFF increases sharply (and in fact goes to infinity) 
both at the magnetic pole ($\xi = 0$) and at the PC rim ($\xi = 1$).  
These features, referred to as slot gaps by Arons, form at the pole 
and equator due to different effects.  The gap at the pole occurs
because the radius of curvature of the magnetic field lines is infinite
there, causing the pair attenuation length to go to infinity.  The
electrons keep accelerating to high altitudes and thus reach high
energies.  At the PC rim, the boundary condition on the potential 
($\Phi = 0$), screens $E_{\parallel}$ close to the rim, preventing 
electron acceleration to high enough energy to produce any pairs.
Thus, $\gamma_{\rm max}$ also goes to zero at the rim.  
Figure \ref{fig:pffxb} shows the dependence of the PFF height and 
acceleration voltage on surface magnetic field strength $B$. While the height
of the PFF decreases with increasing $B$, due both to an increase in 
$E_{\parallel}$ and a decrease in pair attenuation length, 
$\gamma_{\rm max}$ is roughly constant even though $B$ varies over two 
decades.  This is because the longer total acceleration distance for 
lower field strengths compensates for the lower rate of acceleration.

Figure \ref{fig:pffsx} shows examples of ICS-controlled PFFs near the 
NS surface.  We find that both the height of the PFF and the
acceleration voltage is much lower than for the CR-controlled PFFs.  
This is due to the higher efficiency of ICS in producing photons of 
pair-producing energy.  ZQLH97 obtained a similar result 
for ICS-controlled PFFs using the electric field of RS95. 
Electrons having Lorentz factors of only
$10^5-10^6$, depending on $\xi$, are capable of radiating
photons of about $10\%$ to almost $100\%$ of their energy, and these
photons will 
produce pairs. Again, the maximum Lorentz factor, 
$\gamma _{\rm max}$, is a very weak
function of field strength.  The ICS-controlled PFF height, however, 
behaves very differently near the magnetic pole. Instead of forming a 
slot gap, like the CR-controlled PFFs, the height of the
PFF actually decreases near the pole.  ICS photons are radiated at
much larger angles to the field and thus have shorter pair 
attenuation lengths than curvature photons. 

The above determinations of the electron PFF are not fully
self-consistent in that they have neglected pairs produced by the
positrons that slow down to screen $E_{\parallel}$ and accelerate
down toward the stellar surface.  Even if the number of returning 
positrons is small compared to the number of primary electrons, the 
multiplicity of the downward cascades is high (DH96), due to the increasing field strength.  From our simulations of downward going cascades using
the code of DH96, we find that the multiplicity
of a positron accelerated toward the PC from a height of one stellar 
radius at the PC rim is $M_p \sim 10^4 - 10^5$. If the fraction of 
positrons that are accelerated downward from the electron PFF is 
$f_p = \lambda_+ /\lambda_- \sim 3 \times 10^{-3} - 2 \times 10^{-1}$
(see equations [\ref{lam-}] and [\ref{lam+}]), 
then there are $f_p M_p \sim 30 - 2\times
10^3$ pairs produced by downward cascades for each primary
electron. This would seem to be sufficient to screen $E_{\parallel}$ 
and form a lower PFF. 

To compute the location of the lower PFF, a test positron is started
at the upper (electron) PFF with an energy of $\gamma^+ = 1$.  In 
reality, there is a spread in energies of the pairs created near the 
electron PFF, so that the most energetic ones will decelerate but not 
turn-around, while the less energetic ones will decelerate
before they reach the PFF and accelerate downward with varying initial 
energies. We have thus neglected the detailed kinematics of this turn-around
process.  Equation (2) is then integrated to follow the downward
acceleration of the positron and its energy losses due to both ICS and 
CR, in the same manner as for the electron.  Likewise, the attenuation 
lengths of the photons and the total acceleration length of the
positron are computed.  Although the PFFs of the positrons,
like the electrons, are produced by pairs from the ICS process near
the NS surface in the presence of thermal radiation from a hot PC, we find
that the positron PFF is always above the start of the electron acceleration.
That is, pair front formation by the ICS process is not symmetric for 
upward and downward going particles.  This is because the positrons scatter the
thermal photons at much larger angles than the electrons.  While the 
ICS photons producing the electron PFF scatter in the cyclotron
resonance to an energy of $\sim \gamma B'$, photons
producing the positron PFF scatter above the cyclotron resonance, in the 
KN regime (see Fig. \ref{fig:prim1}), where the 
scattered energy is $\sim \gamma$.  This asymmetery will occur for field
strengths $B' \lsim 0.5$, above which positrons will form PFFs through
resonant ICS.  As discussed further in Section \ref{sec:analres},
stable ICS-controlled double PFFs may exist for high-field pulsars
because the scattered energy for resonant scattering is independent of
incident angle.  Figure \ref{fig:epsmin}
shows the difference between the electron, $\gamma^-_{\rm min}$, and 
positron, $\gamma^+_{\rm min}$, energies that produce scattered
photons of energy $\erg^+_{\rm min}$ and $\erg^-_{\rm min}$ that form 
the minimum PFFs, as a function of $h_0$.  
Since their ICS is more inefficient, the electrons must 
accelerate to much higher energies in order to produce scattered photons 
that pair-produce.  The efficiency of electron scattering, 
$\erg^-_{\rm min}/\gamma^-_{\rm min}$, decreases with height, $h_0$, above
the surface, while the efficiency of positron scattering increases.  
Therefore, the positrons are able to radiate pair-producing photons
after traveling a shorter acceleration path. Since the scattered
energies of the positrons, at a given height, are greater 
that those of the electrons, these photons have shorter pair
attenuation lengths. Thus, the total positron acceleration length, 
$S^-_c = S_a (\gamma^-_{\rm min}) + S_p (\erg^-_{\rm min})$ is less
than the total electron acceleration length, $S^+_c$.  

Figure \ref{fig:pffep} shows total positron and electron acceleration lengths,
$S^+_c$ and $S^-_c$ as a function of the electron
starting height $h_0$, for both ICS- and CR-controlled PFFs.  
In these calculations, we have allowed the self-consistent
PFFs of electrons and positrons to form with only one of the radiation 
mechanisms operating. When only ICS produces pairs, $S^+_c$ is always 
significantly less than $S^-_c$ and the difference increases with
height.  However, when ICS is
suppressed and only CR is allowed to operate, $S^+_c$ and $S^-_c$
are equal.  That is because CR for electrons and positrons is the same, 
the only difference being 
in radius of curvature between the radiation points, which is negligible
since $S^-_c \ll R$.      
While ICS cannot provide self-consistent solutions to the double PFFs, 
CR can.  We suggest that stable, self-consistent double PFFs can only exist
when they are formed by CR.  Since ICS will dominate near the 
surfaces of NSs radiating thermal soft X-rays, evidence of 
which has been observed in many pulsars, stable PFFs can only form at
a height above the surface where CR becomes dominant.  We take a rough
estimate for the height of CR-control of the electron and positron PFFs
to be where CR energy loss of the electrons dominates
over ICS energy loss. If electrons of energy $\gamma^-_{\min}$ radiate 
more curvature photons than ICS photons, then the ICS PFF will
disappear, and the electrons will continue accelerating until
a curvature PFF is established.  When the positrons accelerating
downward from the electron curvature PFF also establish curvature 
photon PFFs, then stable acceleration can occur.  In reality, the PFFs
will switch from ICS to CR control when the number of pairs produced by 
ICS photons becomes too small to screen $E_{\parallel}$.  But to determine the 
number of pairs necessary for establishing a lower PFF, we would need 
to know the number of returning positrons.  Figure
\ref{fig:gdot} shows an example of the electron energy loss rates
due to ICS and CR at energy $\gamma^-_{\min}$ as a function of height
of the lower PFF.  At low altitudes, ICS losses dominate by many
orders of magnitude, but decrease with $h_0$ mostly due to the
decrease in the thermal photon density (cf. equation [\ref{n_ph}).  The
energy $\gamma^-_{\min}$ of electrons that are forming
the ICS PFF increase with altitude due to the decreasing efficiency of
ICS. Their CR losses, which are a strong function of energy, 
therefore increase.  The height at which electron ICS and CR  
losses are equal, which roughly set the location 
of stable acceleration, will depend on the PC thermal temperature and 
radius (cf. equation [\ref{h_0}]).

We have computed the height of the stable acceleration zones where 
CR losses dominate control of PFFs for various pulsar
parameters.  As shown in Figure \ref{fig:pffh0},
the location of the lower (positron) PFFs $h_0$ are at higher 
altitudes for higher surface magnetic field strengths, because the 
ICS losses are proportional to $B^2$ for resonant scattering.  The
height of the lower PFF decreases for increasing pulsar rotation
period, $P$, because at a given colatitude $\xi$ the radius of 
curvature increases, but the electron energy $\gamma^-_{\min}$
increases so that CR can dominate at a lower altitude.  Figure 
\ref{fig:pffp} shows the width of the CR-controlled 
acceleration zone, $S^-_c$ (i.e. 
the distance between the lower and upper curvature PFFs)
and the maximum voltage drop, $\gamma_{\rm max}$,
as a function of pulsar period and surface field strength.  The
maximum particle energy is again remarkably insensitive to pulsar parameters,
even though the acceleration zone width shows substantial variation.  
The width  $S^-_c$ tends to be smaller for larger fields
because both the acceleration length and the pair attenuation length
are shorter. It has a large increase at long periods, where the pair 
attenuation lengths become large due to the increase in the PC radius 
of curvature.  In fact, there will be a maximum period for which
electron PFFs can form at a given colatitude.

Figure \ref{fig:pffx} shows solutions for the width and maximum 
Lorentz factor of the
self-consistent, stable acceleration zones that are controlled by CR
double PFFs.  The width of these zones, located nearly a stellar radius above 
the surface, are larger than CR-controlled zones at the surface (e.g. 
Fig. \ref{fig:pcacch} and \ref{fig:pffxb}), because the magnetic field 
has fallen from its surface value, so that $E_{\parallel}$ is lower,
and the radius of curvature is larger.  However, the maximum Lorentz factor of
the higher altitude acceleration zones is roughly the same as that
at the surface.  Again, the size of the acceleration path has adjusted 
itself to maintain the same maximum energy, which depends only on the 
geometry of the magnetic field.

\subsection{Analytic Estimates} \label{sec:analres}

It is difficult to derive accurate analytic expressions for the
acceleration zone parameters we have computed numerically, such as height
and width of the stable accleration zone. The form of the electric
field changes as a function of acceleration length (see Appendix), 
starting as a linear function and then saturating at $s \sim
R_{\rm pc}$ to become a constant.  
Consequently, there are no simple expressions for the processes that
apply for altitudes from the surface to several stellar radii.  
Nevertheless, we can characterize the behavior of the solutions
in different regions.  An estimate for $h_0$, the height at which control 
of the PFF formation switches from ICS to CR, can be obtained by finding 
the height at which the
loss rates for CR and ICS of positrons with energy $\gamma_{\rm min}$
are equal:
\be
\dot\gamma_{_{\rm IC}}(\gamma_{\rm min}) = 
\dot\gamma_{_{\rm CR}}(\gamma_{\rm min}).
\ee 
We can obtain an estimate of $\gamma_{\rm min}$ by solving equation
(\ref{Sc}).  For $s \lsim 3 R_{\rm pc}$, we can use the expression for 
$E_{\parallel}$ given in equation (A3) of the Appendix, which is linear in 
$s = zR_E$ and thus gives a quadratic energy increase in particle
energy with $s$. The 
particle acceleration length is then:
\be
S_{\rm a}(\gamma) = (\gamma/A_{\gamma})^{1/2}
\ee
where
\be
A_{\gamma} \simeq \left({3e\over 2mc^2}\right) \left({\Omega R\over
c}\right) {B_0 \over {1-\epsilon }} \left({\kappa\over R_E}\right) 
\left({R\over R_E}\right)^2 \cos\chi .
\ee
Note that we have rescaled equation (A3) to the effective radius $R_E$, at which
the acceleration begins.
The pair attenuation length can be approximated as the mean free path
\be
S_{\rm p}(\erg) \simeq \left({0.2 \rho_c \over B' \erg}\right)
\ee
valid for $B' \lsim 0.1$.  Equation (\ref{Sc}) may then be written,
\be  \label{Sceq}
S_c = {\rm min}\left\{ (\gamma/A_{\gamma})^{1/2} + 
{0.2\rho_c\over B'\erg}\right\},
\ee
where 
\be
\rho_c \simeq {4\over 3}\,R_E^{1/2}\,\left({\Omega\over c}\right)^{1/2}
\ee
and 
\be
B' \simeq \left({B_0\over B_{\rm cr}}\right)\,\left({R_E\over R}\right)^{-3}.
\ee
We wish to estimate $\gamma_{\rm min}$ when ICS produces
the PFFs.  For nearly all pulsar field strengths ($B \lsim 0.5\,B_{\rm cr}$)
the photons which produce the positron PFFs have been blue-shifted above
the resonance and scatter in the KN regime (see Fig. \ref{fig:prim1}).
In this case, we can approximate the scattered photon energy as $\erg 
\sim \gamma$.  Inserting this value of $\erg$ into equation (\ref{Sceq}), 
we then obtain $\gamma_{\rm min}$ giving the minimum $S_c$ by setting 
the derivative of $S_c$ with respect to $\gamma$ to zero:
\be   \label{gmin_KN}
\gamma^{KN}_{\rm min} = \left({0.4\rho_cA_{\gamma}^{1/2}\over B'}\right)^{2/3}
= 3.4 \times 10^5\,B_{12}^{-1/3}\,\left({R_E\over R}\right)
(\cos\chi)^{1/3}.
\ee
where $B_{12}$ is the surface field strength.
Thus, the particle energy required to radiate pair-producing photons increases
with altitude, as was found in our numerical results 
(cf. Fig. \ref{fig:epsmin}). 

We can obtain an estimate for the width of the ICS controlled
acceleration zones by substituting $\gamma^{KN}_{\rm min}$ from equation 
(\ref{gmin_KN}) into equation (\ref{Sceq}) for the minimum value of $S_c$,
\be
S^{KN}_c \simeq 5.3 \times 10^3\,B_{12}^{-2/3}\,
P^{1/2}\,R_6^{1/2}\,\left({R_E\over R}\right)^{5/2}\,(\cos\chi)^{-1/3}
\,{\rm cm}.
\ee
We can estimate the height $h_0$ above the surface at which CR energy 
loss exceeds ICS losses for the downward moving positrons of energy 
$\gamma_{\rm min}$. From our numerical results, we found that at the 
altitudes where CR begins to dominate the particle energy loss, the
ICS is in the K-N regime above the resonance.  We will therefore
equate the curvature loss rate to the K-N loss rate to obtain an
estimate for $h_0$. Substituting $\gamma_{\rm min}$ from equation
(\ref{gmin_KN}) in equation(\ref{gdot_CR}) for the CR energy loss, we have
\be
\dot\gamma_{_{\rm CR}}(\gamma^{KN}_{\rm min}) = 1.6 \times 10^4
\,B_{12}^{-4/3}\,R_6^{-1}\,P^{-1}\left({R_E\over R}\right)^3\,
(\cos\chi)^{4/3}\,{\rm s^{-1}}
\ee
which sharply increases with altitude, $R_E = R + h_0$.  We approximate the K-N
loss rate, modifying Blumenthal \& Gould's (1970) formula for scattering of
isotropic thermal photons of temperature $T_6$, from a hot polar cap of radius $R_T$, with a dilution factor that
accounts for the fall-off in photon density with height $h$:
\be
\dot\gamma_{_{\rm KN}}(\gamma^{KN}_{\rm min}) \simeq {\pi r_0^2\over 6 mc^2}
{(mckT)^2\over 
\hbar^3}\,\ln\left({4\gamma_{\rm min} kT\over mc^2}\right) {R_T^2\over 
h^2+R_T^2} \simeq 7.6 \times 10^8\,{\rm s^{-1}}\, T_6^2\,{R_T^2\over h^2}
\ee
for $R_T \ll h$ and ignoring the slowly varying $\ln$ factor.  
Setting $\dot\gamma_{_{\rm CR}}(\gamma^{KN}_{\rm min}) = 
\dot\gamma_{_{\rm KN}}(\gamma^{KN}_{\rm min})$ and assuming $R_E \sim h_0$, 
\be \label{h_0}
\left({h_0\over R}\right)_{KN} \simeq 8.6\,B_{12}^{4/15}\,P^{1/5}\,R_6^{1/5}
\,T_6^{2/5}\,\left({R_T\over R}\right)^{2/5}(\cos\chi)^{-4/15}.
\ee
Although the above expression involves a number of approximations, it does
seem to roughly reproduce our numerical results.  For example, taking the
parameters $T_6 = 0.5$, $R_6 = 0.8$,
$P = 0.1$, $B_{12} = 4.4$ and $R_T = 0.04\, R$ gives $h_0/R \simeq 1.6$,
fairly close to the value shown in Fig. \ref{fig:pffp}.  

At the altitude $h_0$, where CR losses are dominant and curvature
photons produce the PFFs, the width of the acceleration zone can be 
calculated in a similar manner to that of the ICS controlled
acceleration zones.  Using the CR critical frequency, $\erg_{\rm cr} = 
(3/2)(\hbar/mc) \gamma^3/\rho_c$ in equation (\ref{Sceq}), we can solve 
for the particle energy $\gamma^{CR}_{\rm min}$ which produces the 
minimum acceleration zone width
$S_c$,
\be  \label{gmin_CR}
\gamma^{CR}_{\rm min} = (6 C_p A_{\gamma}^{1/2})^{2/7}=
4.7\times 10^7 B_{12}^{-1/7}P^{1/7}\left({ {R_{\rm E}}\over R}
\right) ^{4/7} R_6^{1/7} (\cos \chi )^{1/7},
\ee
where 
\be
C_p = \left({0.2\rho_c\over B'\epsilon_{\rm cr}}\right).
\ee
Then substituting the expression for $\gamma^{CR}_{\rm min}$ back into 
equation (\ref{Sceq}) for $S_c$, we have
\be \label{Sc_CR}
S^{CR}_c \simeq 4.8 \times 10^4\,B_{12}^{-4/7}\,
P^{4/7}\,R_6^{4/7}\,\left({R_E\over R}\right)^{16/7}\,
(\cos\chi)^{-3/7} \,{\rm cm}.
\ee
The expression for $\gamma^{CR}_{\rm min}$ above is also a very good 
estimate of $\gamma^{CR}_{\rm max}$, the maximum Lorentz factor of the
CR-controlled accleration zone, because $S_p \ll S_a$ for CR photons 
(i.e. the PFF is very close to where the first pairs are produced).

For very high field strengths ($B' \gsim 0.5$) both electron and positron
ICS operate in the resonant scattering regime. 
The incident soft photons scatter in the cyclotron resonance and their 
scattered energies are approximately $\erg_{\rm s} \sim \gamma B'$. 
In the Thomson limit of resonant scattering, the energy of the scattered
photons is independent of incident photon angle.  This means that $S_c^+ =
S_c^-$, PFF$^+$ will coincide with $h_0$, and stable ICS-controlled 
acceleration zones are thus possible for high-field pulsars.
However, in the relatively high fields where resonant ICS controls
the positron PFFs, relativistic effects on the resonant scattering cross
section as well as photon splitting will become important (as will be
discussed in Section \ref{sec:Dis}) and our calculation of PFFs is
incomplete.

\section{Discussion} \label{sec:Dis}

In this paper, we have investigated the effect of cascades from
downward-accelerated positrons on the electrodynamics of the PC particle
acceleration.  We find that when ICS produces pairs in the acceleration zone,
the positron cascades may screen the accelerating electric field and disrupt
particle acceleration near the NS surface.  Thus, if lower PFFs can 
develop, the picture of steady particle acceleration from the PC surface must
undergo major
revision.  We suggest that a stable acceleration zone may exist at an altitude 
of about one stellar radius above the PC, in which pairs from CR
limit the electrostatic acceleration of primaries by screening the 
electric field near the upper PFF.

The calculations presented here are only a first attempt to describe the
physics of what is a very complicated process.  We have made many assumptions
and approximations to obtain our results.  While we believe that the gross
qualitative results of our study are correct, there are a number of aspects
which should be treated more accurately to achieve more solid quantitative
results.  We have assumed that the screening of the $E_{\parallel}$ occurs over
the short distance determined by the upper boundary condition of Poisson's
equation, not by the dynamics of the pair screening.  The
details of the pair screening also determine the fraction of pairs that return
to the PC and we have briefly outlined in Section (\ref{sec:pfraction}) 
how such a calculation 
can be done within the electrodynamic framework set up in this paper.  A
determination of the returning positron fraction would answer some interesting
questions, such as whether PC heating is important relative to cooling
in setting the PC temperature and if so, whether there is a feedback loop 
between the formation of the upper PFF and the PC heating.  The characteristics
of the downward cascades need further study and modeling.  While we have done
preliminary modeling of these cascades near the PC rim, there will be significant
variation in pair yields with magnetic colatitude.  Near the magnetic pole, 
the pair multiplicity will drop, allowing the lower PFF height to decrease or
even disappear.

We have found that the nature of the PC acceleration depends strongly on the
characteristics of the radiation from the hot PC.
Thus, to solidify the quantitative aspects of our results, it is important to
treat the thermal PC radiation and the ICS process as accurately as possible.
For example, we have assumed in this paper, that the thermal radiation is
uniformly emitted over a PC of size $R_T = 3\theta_0$ with an 
isotropic flux distribution.  However, studies of thermal radiation 
propagating through a strongly magnetized NS atmosphere 
(Pavlov et al. 1994) will not be isotropic, due to the
anisotropy of the magnetized scattering cross section.  The expected
radiation pattern consists of a pencil component, beamed along the 
magnetic field and a fan component perpendicular to the field.  
Such a beam pattern has been found to be consistent with observed 
thermal X-rays pulse fractions and pulse profiles for several pulsars 
(Shibanov et al. 1995, HM98).  In addition, our treatment of
ICS uses a combination of the (non-relativistic) magnetized cross 
section in the 
Thomson limit for scattering near the fundamental cyclotron resonance, and the 
(non-magnetic) Klein-Nishina cross section to describe relativistic effects 
above the resonance.  
While this hybrid treatment is somewhat inaccurate for $B \gsim 0.1\,B_{\rm cr}$ 
and therefore not completely satisfactory, the fully relativistic QED magnetic
scattering cross section (e.g. DH86) is too complicated for 
use in this type of calculation.  In particular, the quantization of the electron momentum perpendicular to the field limits the number of Landau states contributing 
to the cross section for each incident photon energy.  The magnetic QED cross 
section for scattering just above the fundamental cyclotron resonance, and
therefore the loss rate, could therefore be substantially lower than the 
Klein-Nishina cross section we have used here.  
Unfortunately, there exists no simplified, 
approximate expression for the QED scattering cross section, which smoothly
bridges, and allows a unified treatment of, the relativistic resonant 
and non-resonant regimes of ICS. 

A physical process which we have neglected in this study, photon splitting,
is not expected to be significant for the magnetic fields we have considered,
but will be very important for pulsars having surface 
$B_0 \gsim 0.5\,B_{\rm cr}$.
Photon splitting, a third-order QED process in which one photon splits 
into two, operates
only in very high magnetic fields and competes as an attenuation process with
one-photon pair production (Harding et al. 1997) because it can occur below 
pair threshold.  The implications of
photon splitting for PC PFFs is profound.  Pulsars having 
$B_0 \gsim 0.5\,B_{\rm cr}$ will produce fewer pairs, especially near 
the surface, so that the cascades from
downward-moving positrons may not produce a lower PFF.  This is about the same 
field strength where we found that double PFFs controlled by resonant ICS 
at the NS surface become possible.  Thus, the
stable, CR-controlled double PFF structure we have studied in this paper, 
which we found to move
to higher altitudes with increasing surface field strength, will eventually
collapse back to the surface at very high $B_0$.  A single PFF will then form,
controlled by ICS, and the acceleration zone will have characteristics similar to
that shown in Figure \ref{fig:pffsx}.  At extremely high surface fields, 
$B_0 \gsim 1.0\,B_{\rm cr}$ photon splitting will suppress pair creation completely 
at this single PFF, and the pulsar may be radio quiet (Baring \& Harding 1998).
However, the particle acceleration in these pulsars will operate very efficiently,
free of any screening of $E_{\parallel}$, so they are expected to be observable
at high energies.  Bound-state pair production, where photons convert to 
positronium just below pair threshold rather than converting to free pairs 
(Usov \& Melrose 1995), may also come into play at higher field strengths.  

Some of the main results that we have presented in this paper have some
important implications for pulsar high-energy emission.  One of these results
is the insensitivity of the acceleration voltage (maximum particle energy)
$\gamma_{\rm max}$ to any pulsar parameters such as period, surface magnetic
field strength, obliquity (except for nearly orthogonal rotators), and
even height. This
result showed up many times in the course of our calculations (cf. Figs.
\ref{fig:pcacch}, \ref{fig:pffxb}, \ref{fig:pffsx}, \ref{fig:pffp} and 
\ref{fig:pffx}, and equation [\ref{gmin_CR}]), and seems
to be a robust characteristic of this type of PC acceleration model.  The
maximum particle acceleration energy varies only within each pulsar, as a 
function of magnetic colatitude.  This energy, between $5 \times 10^{12}$ eV and 
$5 \times 10^{13}$ eV for CR-controlled acceleration zones, is about two to
three times higher than that without frame-dragging (e.g. A83), and is consistent 
with the primary particle energy
required in CR-initiated PC cascade models of $\gamma$-ray pulsars (e.g. 
DH96).  The acceleration energy is very high near the magnetic poles, where the
radius of curvature goes to infinity, allowing for the possibility of a narrowly
beamed, hard component in $\gamma$-ray pulses.  However, the radiation power 
emitted by these high-energy particles will be small, because the curvature 
radiation loss rate is proportional to $\rho_c^{-2}$. 
The insensitivity of $\gamma_{\rm max}$ to pulsar parameters implies that 
the primary particles in all pulsars are accelerated to the same energy, and
that the luminosity of the high-energy emission should depend only of the
flux of primary particles.  This is consistent with trends in the observed 
$\gamma$-ray pulsar luminosities (Thompson et al. 1997).  The $\gamma$-rays
will originate within a stellar radius of the upper stable PFF, at height
$h_c = h_0 + S^-_c$.  From our results of Figs. \ref{fig:pffh0} and \ref{fig:pffp}, and
equations (\ref{h_0}) and (\ref{Sc_CR}), the height of the $\gamma$-ray production 
increases with period, roughly as $P^{1/2}$.  The standard PC
half-angle at height $R_{\gamma } = R + h_c$ will be 
$\theta_c \simeq (\Omega R_{\gamma }/c)^{1/2} 
\propto P^{-1/4}$.  Thus, the $\gamma$-ray emission solid angle $\Omega_{\gamma}$, 
which is expected to be, $\Omega_{\gamma} \simeq 2\pi [1-\cos(3\theta_c /2)]$, 
will be very weakly dependent on period and field strength. 

Our conclusion that stable acceleration may occur in most pulsars at some
altitude above the surface will have consequences not only for high-energy 
emission, but for radio emission as well.  If electron-positron pairs are
necessary for coherent radio emission, then the dependence of the PFF altitude
on pulsar parameters should be taken into account when determining the 
radio pulsar ``death line", the line on the period-period derivative diagram
beyond which pulsars are incapable of producing pairs.  Our
calulcations in this paper suggest that pulsars with long periods do 
not produce PFFs, and that there will be a ``death line" at periods below
where there are observed radio pulsars.  But this is a 
long-standing problem of PC acceleration, most recently discussed by 
Arons (1998), and one that should be addressed in future studies.

Our principal findings can be summarized as follows.

\begin{enumerate}

\item Lower PFFs may form by positrons returning to the NS surface 
from the upper PFF.

\item Pair creation by the ICS process dominates near the stellar surface,
but is not symmetric
for upward and downward going particles, so that stable, double PFF formation
is very unlikely.

\item Stable, self-consistent double PFFs can only exist when they 
are formed by CR. They can only form at a height above the surface
where CR becomes dominant.

\item The maximum particle energy is insensitive to any pulsar
parameters such as period, surface magnetic field strength, obliquity, 
even height, and is sufficient to power $\gamma$-ray pulsars.

\end{enumerate} 

The main conclusion of this paper is that the cascades from
positrons returning to the PC may have a significant effect on the primary
particle acceleration in pulsars and should not be neglected.  It is possible 
that lower PFFs do not form for all pulsars, and may not form 
over the entire PC.  A detailed study of the screening of the
accelerating electric field by the returning positron cascades is beyond the
scope of this paper, but will ultimately be necessary to understand 
PC acceleration. These studies are needed to address 
the questions of the returning positron fraction and the 
multiplicity of downward cascades.  If we can show that the returning 
positrons do not screen $E_{\parallel}$, then the present assumption 
of acceleration right from the NS surface is valid.  But if these 
studies show that screening at a lower PFF is effective, then the 
possibility of acceleration above the NS surface must be incorporated 
in PC models.

We thank the referee Bing Zhang for his very careful review and 
insightful comments.  
We are also grateful to Joe Daugherty, for help in the downward cascade 
simulations, to Steve Sturner for discussions on inverse-Compton
scattering and Matthew Baring for comments on the manuscript.

\clearpage

\section*{APPENDIX A. APPROXIMATE EXPRESSIONS FOR THE ACCELERATING 
ELECTRIC FIELD}

Here we present the explicit expressions for the electric field
component parallel to the magnetic field for the various ranges of 
altitude.  These expressions have not been rescaled to allow for acceleration
starting at an effective radius above the stellar surface, so that $z$
is the altitude above the surface.\\

\noindent ~~$\bf S_{\rm c} \ll r_{pc}$, and 
$\bf 0.1\,r_{pc}(r_{pc}/R) \ll s \leq S_{\rm c}$:
$$
E_{\parallel} \simeq -3 {{\Omega R}\over c} {B_0 \over
{1-\epsilon }} \left( 1 - {z \over z_c}\right) z z_c \left[ \kappa \cos
\chi + {1\over 2} \Theta_0 \xi H(1) \delta (1) \sin \chi \cos \phi \right],
\eqno (A1)
$$
\noindent~~$\bf S_{\rm c} \ll r_{pc}(r_{pc}/R)$, and $\bf s \ll S_{\rm c}$:
$$
E_{\parallel} \simeq - {3 \over 2} \left( {{\Omega R}\over c} \right) ^2 
{B_0 \over f(1)} (1 - \xi ^2) z \left[ \kappa \cos \chi + {1\over 4} 
\Theta_0 \xi H(1) \delta (1) \sin \chi \cos \phi \right],
\eqno (A2)
$$
\noindent~~$\bf S_{\rm c} < r_{pc}/3$, and $\bf s \leq S_{\rm c}$:
$$
E_{\parallel} \simeq -3 {{\Omega R}\over c} {B_0 \over
{1-\epsilon }} \left( 1 - {z \over z_c}\right) z \left[ \kappa \cos
\chi + {1\over 2} \Theta_0 \xi H(1) \delta (1) \sin \chi \cos \phi \right],
\eqno (A3)
$$
\noindent~~$\bf s \gsim r_{pc}/3$ and $\bf s\leq S_{\rm c}$:
\begin{eqnarray}
E_{\parallel} & \simeq & - {3\over 2} \left( {{\Omega R}\over c} \right) ^2 
{B_0 \over f(1)} \left\{ \kappa {1\over \eta^4}\,\left[(1-\xi^2) - 
\left({\eta\over \eta_c}\right)^3 \sum_{i=1}^{\infty}\,{8 J_0(k_i\xi)
\over k_i^3\,J_1(k_i)}\,e^{-\gamma_i(\eta_c)(\eta_c-\eta)}\right]
\cos\chi \right. \nonumber \\
&& + {1\over 4} \Theta_0 \left[ f(1)\over f(\eta ) \right] ^{1/2}\,
F(\eta ) {1 \over {\eta ^{1/2}}} \left[ \xi(1-\xi^2) - {\eta_c \over \eta } 
{{F(\eta_c)} \over {F(\eta )}} \right. \nonumber \\
&& \left. \left. \sum_{i=1}^{\infty}\,{16 J_1(\tilde k_i\xi)\over 
\tilde k_i^3\,J_2(\tilde k_i)}\,
e^{-\tilde\gamma_i(\eta_c)(\eta_c-\eta)}\right]
\sin\chi\cos\phi\right\}, \nonumber 
\end{eqnarray}
$$\eqno(A4)$$\\
which translates into
$$
E_{\parallel} \simeq - {3 \over 2} \left( {{\Omega R}\over c}\right) ^2 {B_0 \over
f(1)} \left( 1 - \xi ^2 \right) \left[ \kappa \cos \chi + {1\over 4} 
\Theta_0 \xi H(1) \delta (1) \sin \chi \cos \phi \right],
\eqno (A5)
$$
when $\bf s\ll S_{\rm c}$.\\
Here 
\begin{eqnarray}
F(\eta) & = & - {2 \over \eta } \left( \epsilon - {{2 \kappa }\over 
{\eta ^2}}\right) + {3 \over {(1 - \epsilon /\eta )f}} 
\left[ {1\over \eta }\left( \epsilon - {\kappa \over
{\eta ^2}}\right) \right. \nonumber \\
&& \left. - {1 \over {(1 - \epsilon /\eta ) }} 
\left( {4\over 3} - {\epsilon \over \eta } - {3 \over {2f}} \right) 
\left( 1 - {3\over 2} {\epsilon \over \eta } + {\kappa \over {2
\eta ^3}} \right) \right], \nonumber
\end{eqnarray}
$$
\Theta_0 = \left[ {{\Omega R}\over c} {1 \over f(1)} \right] ^{1/2}.
$$
In the above expressions $r_{\rm pc} \approx \Theta _0~R$ is the PC 
radius, $z_c \equiv S_{\rm c}/R$ is the dimensionless altitude of the
upper boundary (PFF), and all other quantities are scaled with the 
true stellar radius. Note also that $H(1)\delta(1)\approx 1$.

\clearpage
\vskip -1.0 in
\figureout{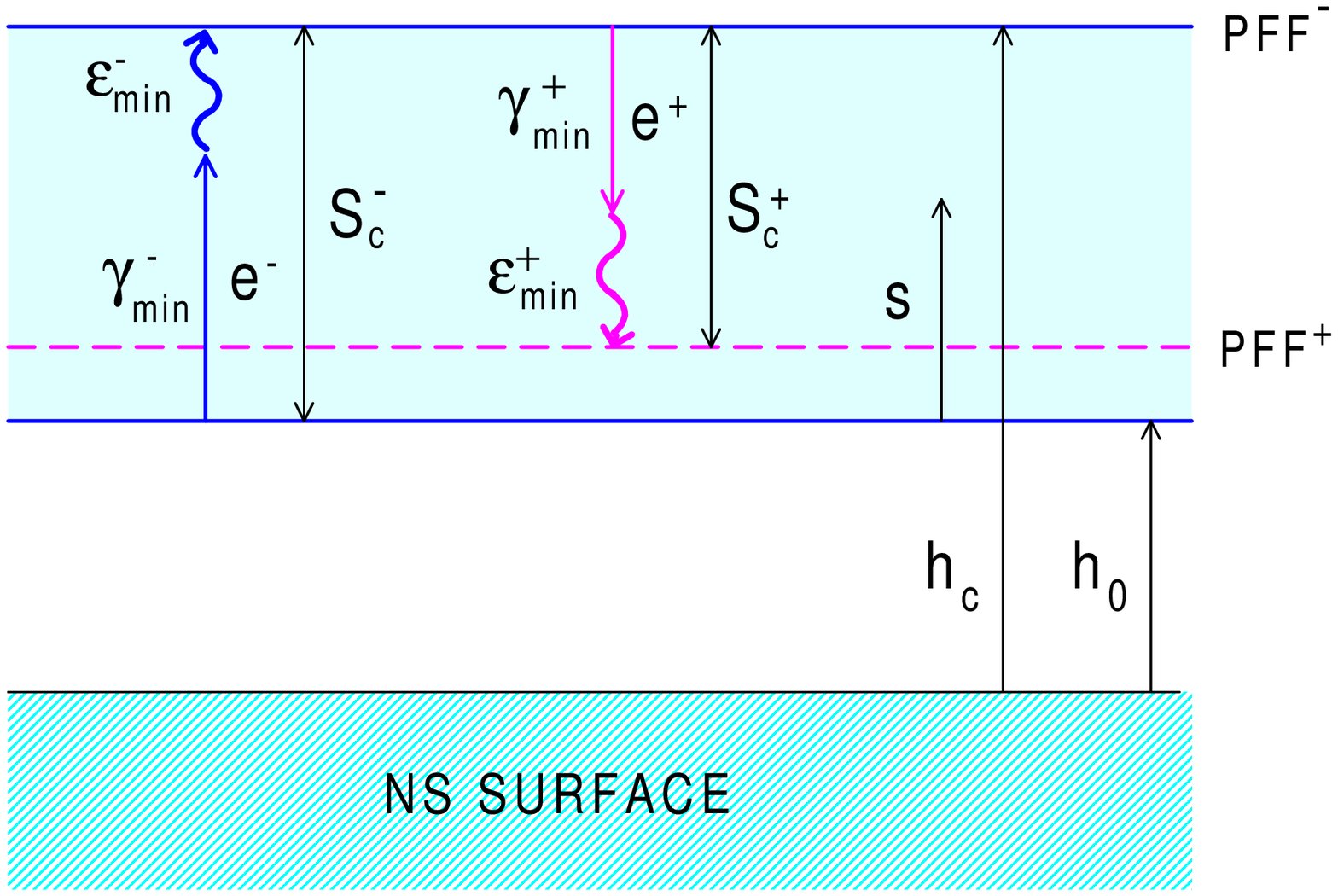}{0}{
Schematic illustration of the double pair formation fronts (PFF) produced
by electrons and positrons at height $h_0$ above a pulsar PC.  
Parallel electric field accelerates particles in the shaded zone, up
to the electron PFF$^-$ at height $h_c$, 
and is screened (${\bf E \cdot B} = 0$) 
everywhere else.  Upward accelerating electrons of energy
$\gamma^-_{\rm min}$ radiate photons of energy $\erg^-_{\rm min}$
that can produce the first electron-positron pairs to form the PFF$^-$.  
Downward accelerating positrons of energy $\gamma^+_{\rm min}$ radiate photons 
of energy $\erg^-_{\rm min}$ that produce pairs which may screen the electric
field at PFF$^+$.  This configuration shows double ICS-controlled PFFs,
which are not stable because the positron PFF$^+$ lies about $h_0$. 
\label{fig:pff} }        

\figureout{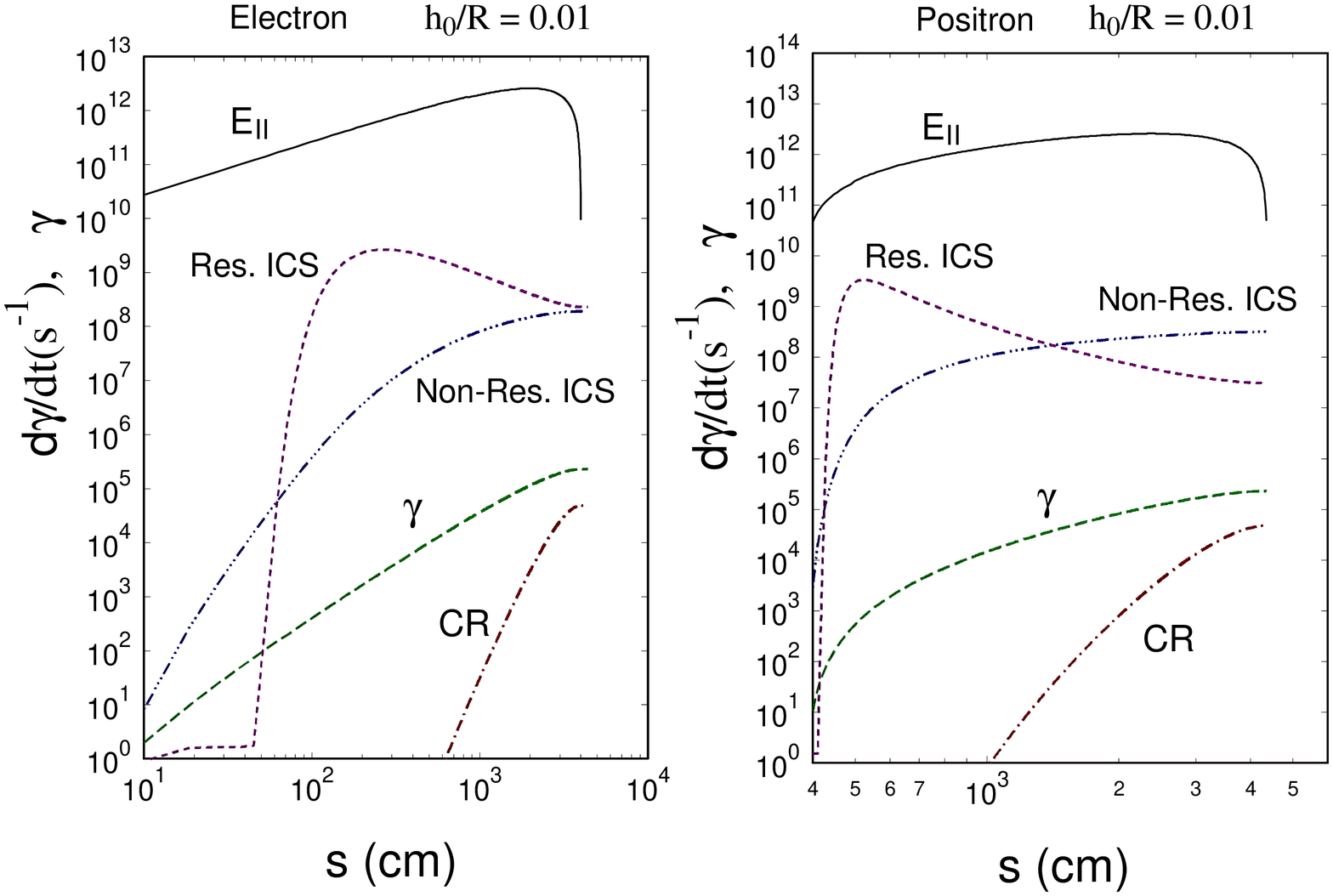}{0}{
Parallel electric field $E_{\parallel}$, particle energy $\gamma$ and energy
loss rates $\dot\gamma$ due to non-resonant and resonant ICS and CR as a function
of acceleration length $s$ for ICS-controlled PFFs of electrons (left panel) and 
positrons (right panel). Here, $h_0$ is the height of the lower PFF (start of 
acceleration) above the surface, $T_6 = 0.5$, $P = 0.1$ s, $B = 0.1\,B_{\rm cr}$,
$\chi = 0.2$ (radians), and $\xi = 0.7$.
   \label{fig:prim1} }    

\psfig{figure=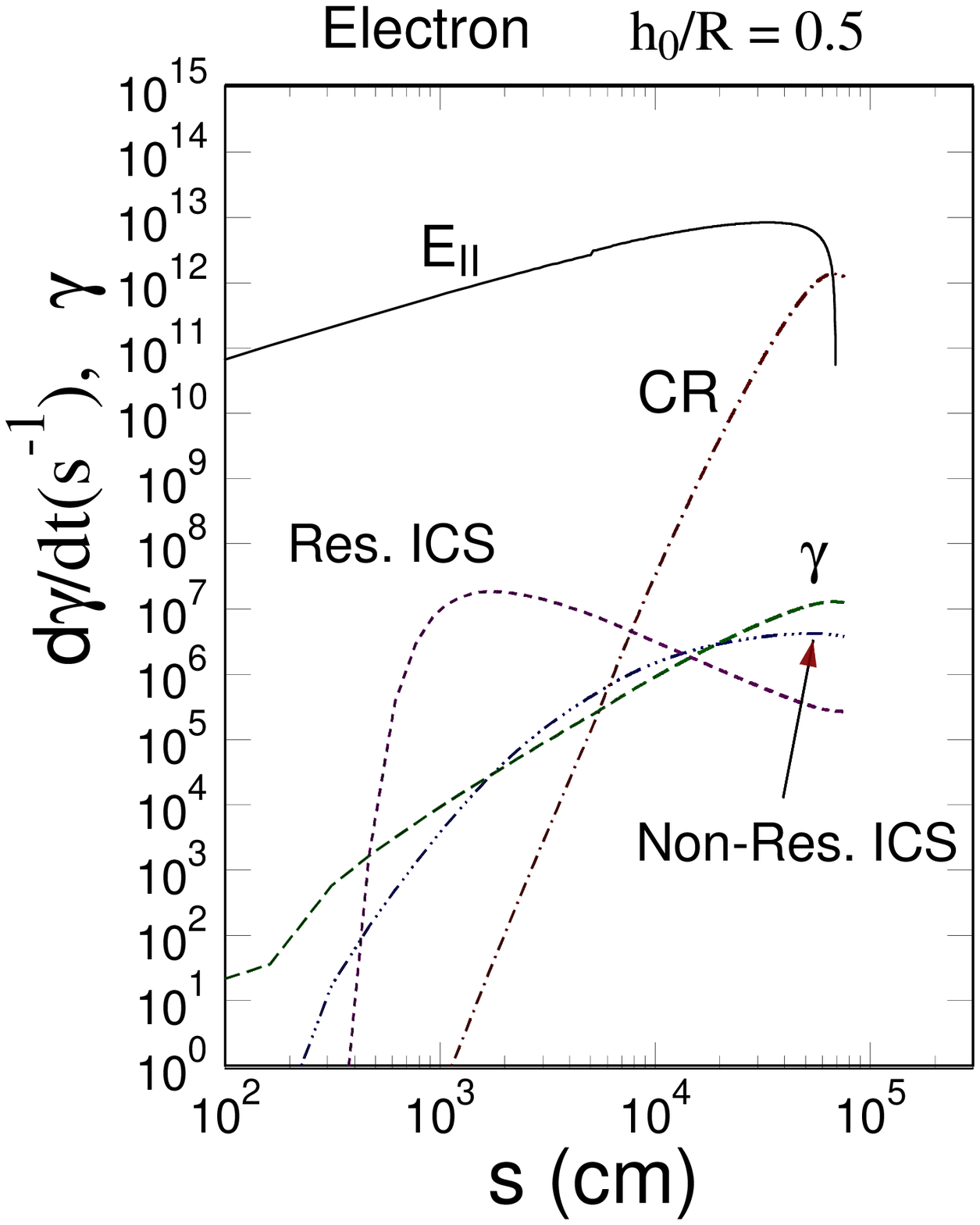,width=6in,angle=0} 
\figcaption{
Same as Fig. \ref{fig:prim1}, but for the electron CR-controlled PFF at height
$h_0 = 0.5 R$.
   \label{fig:prim2} }    

\figureout{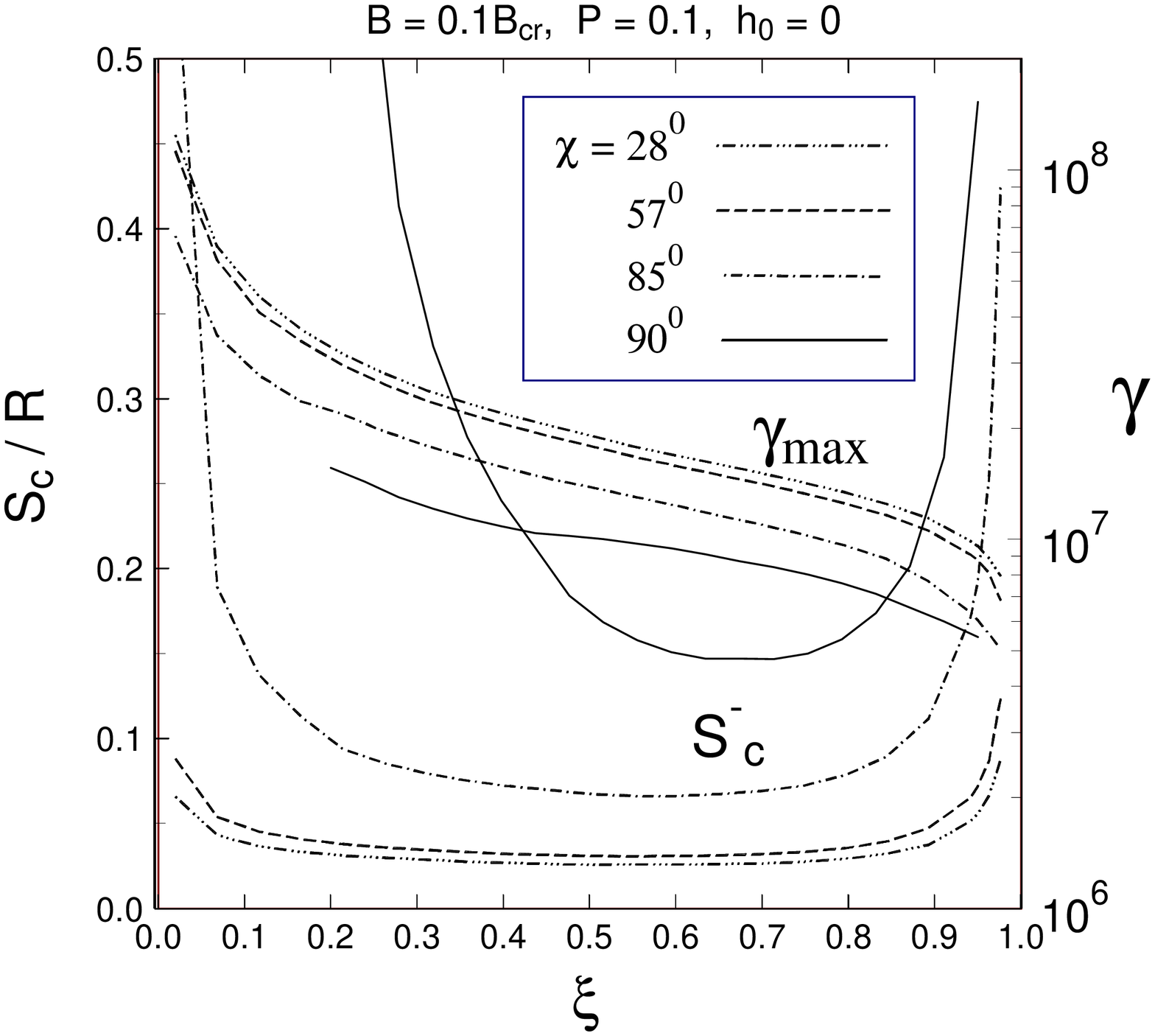}{0}{
Profiles of the acceleration voltage (maximum electron Lorentz factor) $\gamma_{\rm max}$
and width, $S^-_c$, of the acceleration zone formed by a CR-controlled PFF at the 
NS surface, neglecting losses and pairs from ICS, as a function of magnetic 
colatitude scaled to the PC half angle $\xi = \theta/\theta_0$, for different
obliquities $\chi$.
   \label{fig:pcacch} }        

\figureout{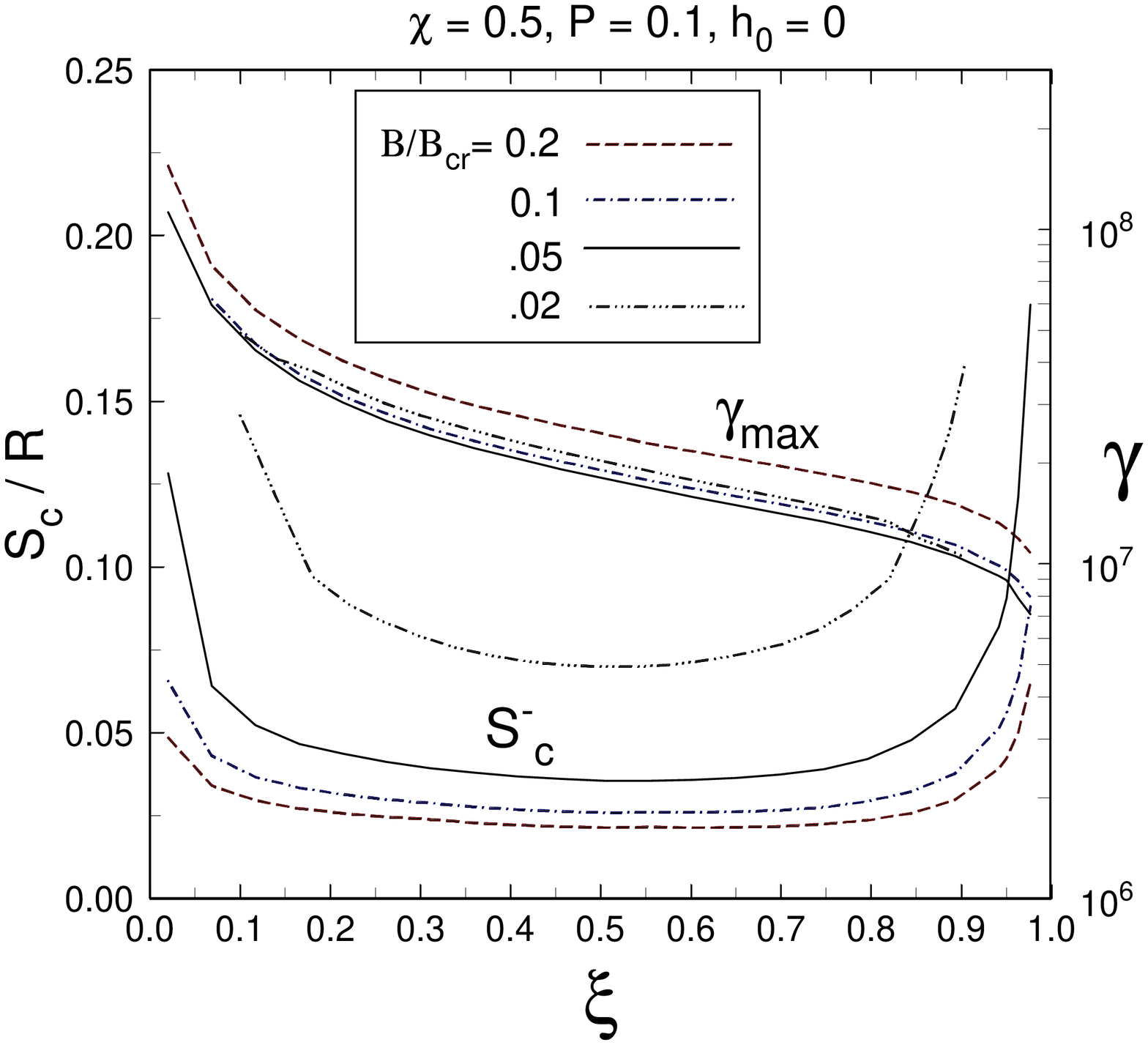}{0}{
Same as Fig. \ref{fig:pcacch} but for different values of the surface magnetic 
field strength.  In Figs. 5-10, $\chi$ is in radians.
   \label{fig:pffxb} }        

\figureout{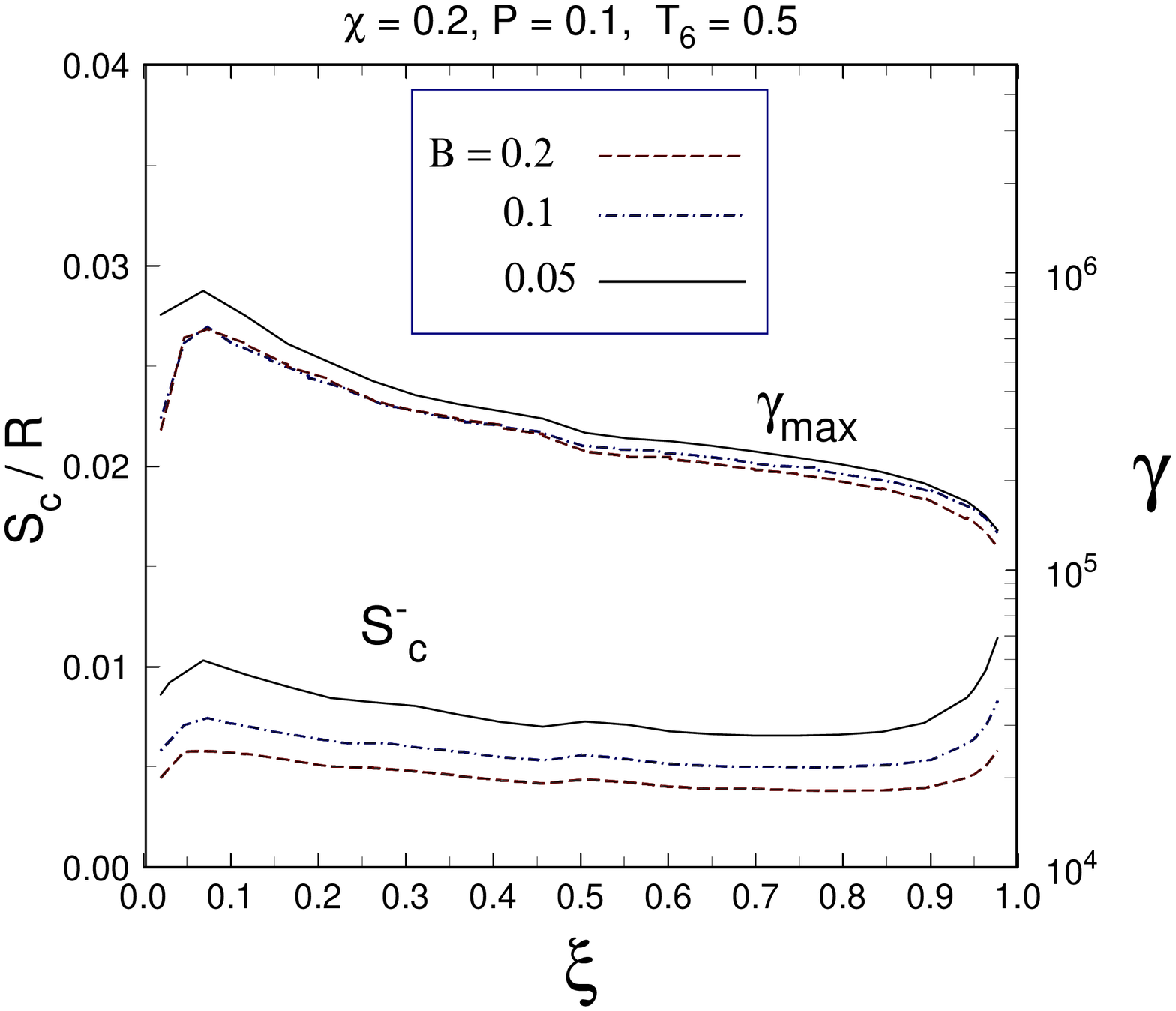}{0}{
Same as Fig. \ref{fig:pffxb} but for ICS-controlled PFFs at the NS surface.
   \label{fig:pffsx} }        

\figureout{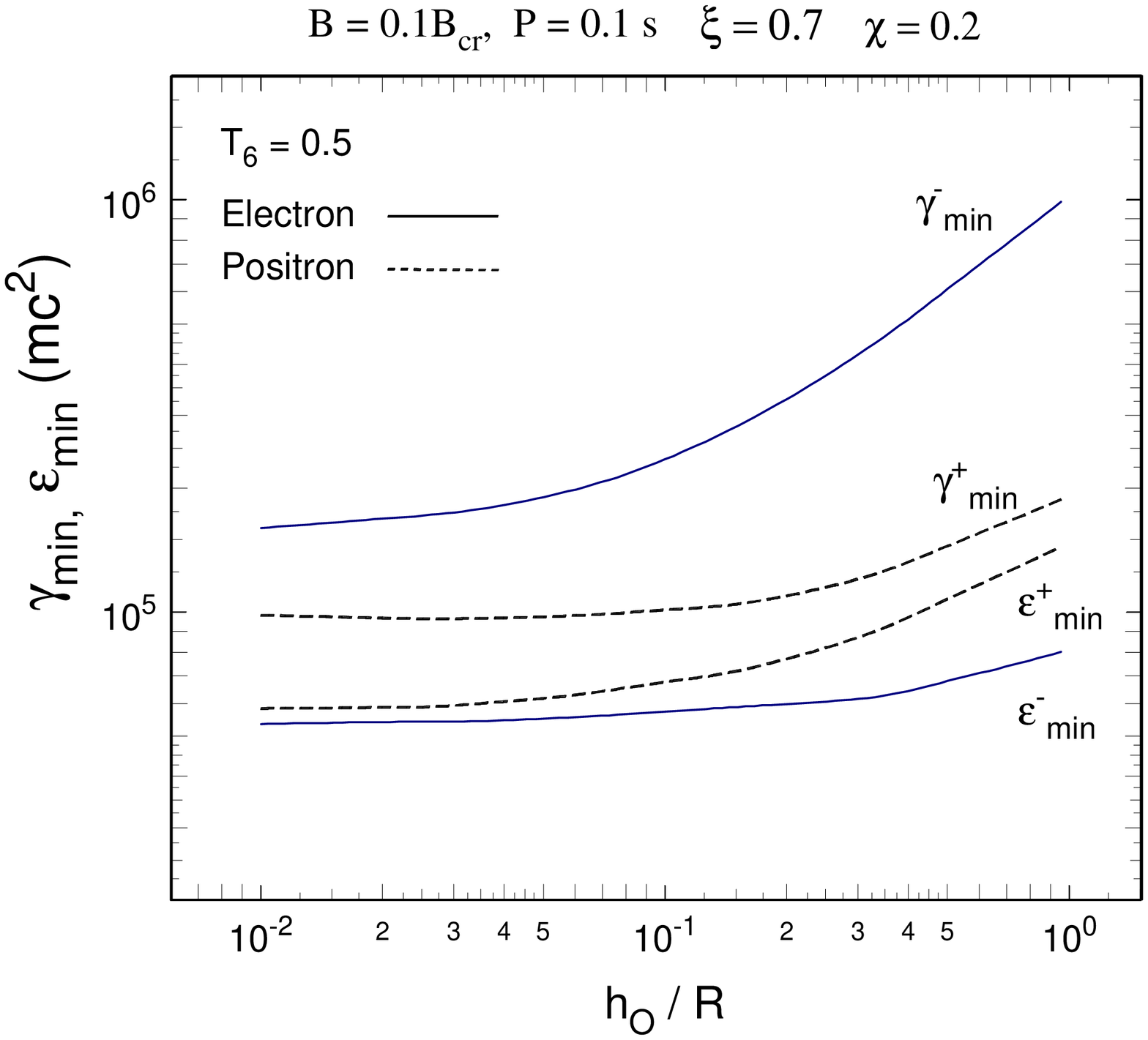}{0}{
Minimum Lorentz factors of electrons, $\gamma^-_{\rm min}$, and positrons,
$\gamma^+_{\rm min}$ that radiate pair-producing
ICS photons of energy $\erg^-_{\rm min}$ and $\erg^+_{\rm min}$, as a function
of acceleration starting height $h_0$.  
   \label{fig:epsmin} }        

\figureout{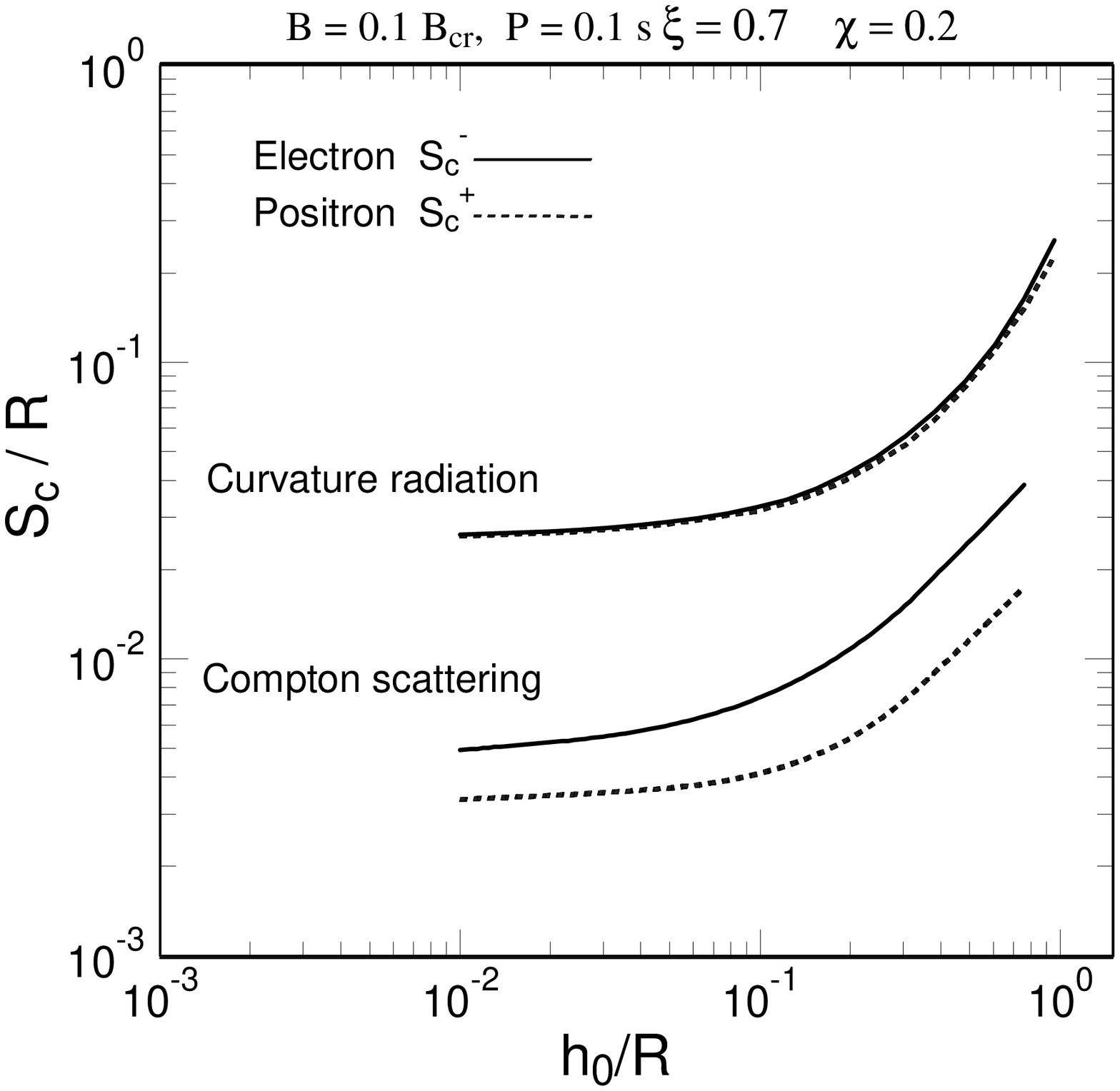}{0}{
Total positron ($S^+_c$) and electron ($S^-_c$) acceleration lengths for CR-controlled
and ICS-controlled PFFs as a function of acceleration starting height $h_0$.  
   \label{fig:pffep} }        

\figureout{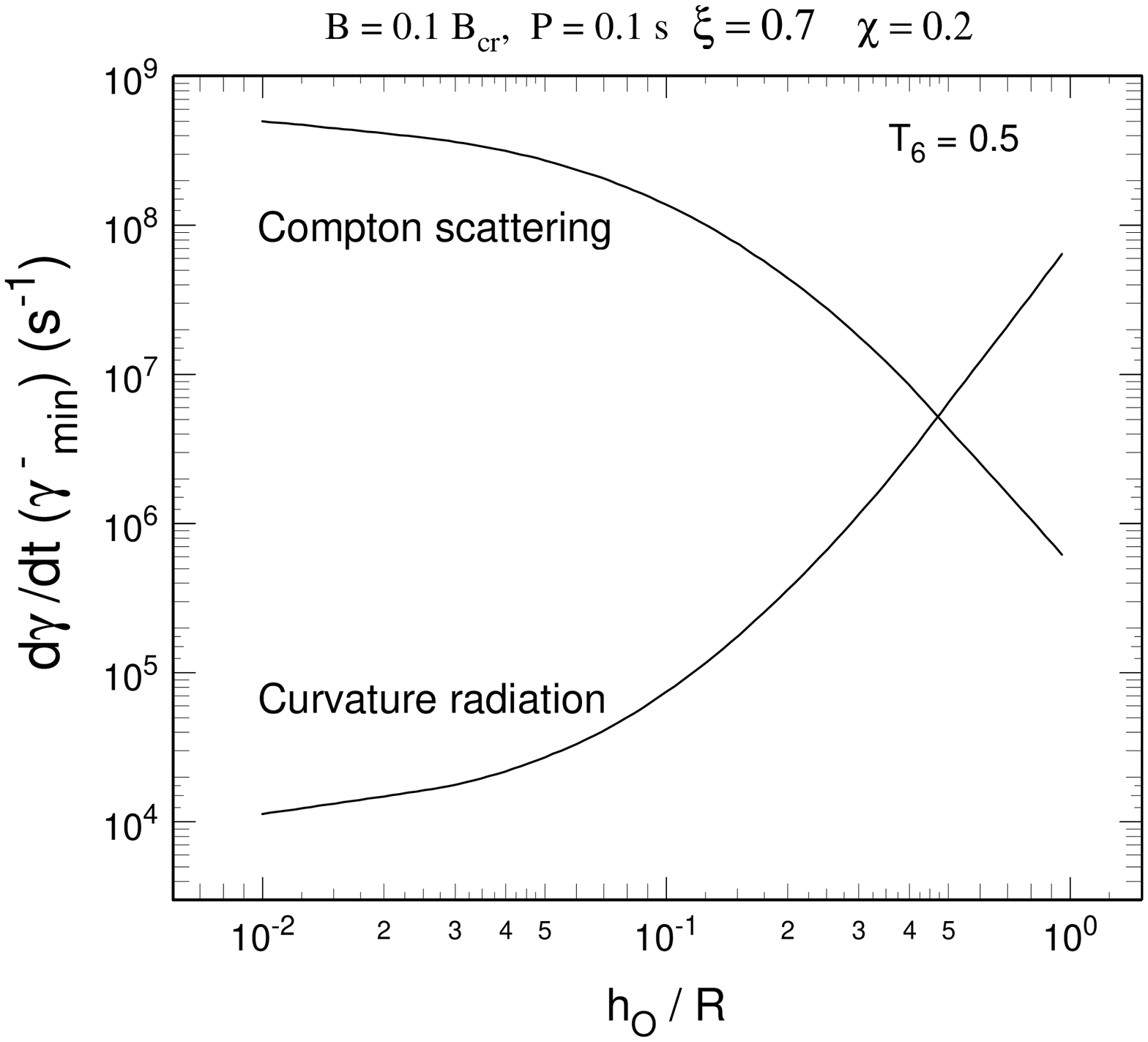}{0}{
Electron energy loss rates for CR and ICS 
at Lorentz factors $\gamma^-_{\rm min}$, as a function of
acceleration starting height $h_0$. 
   \label{fig:gdot} }        

\figureout{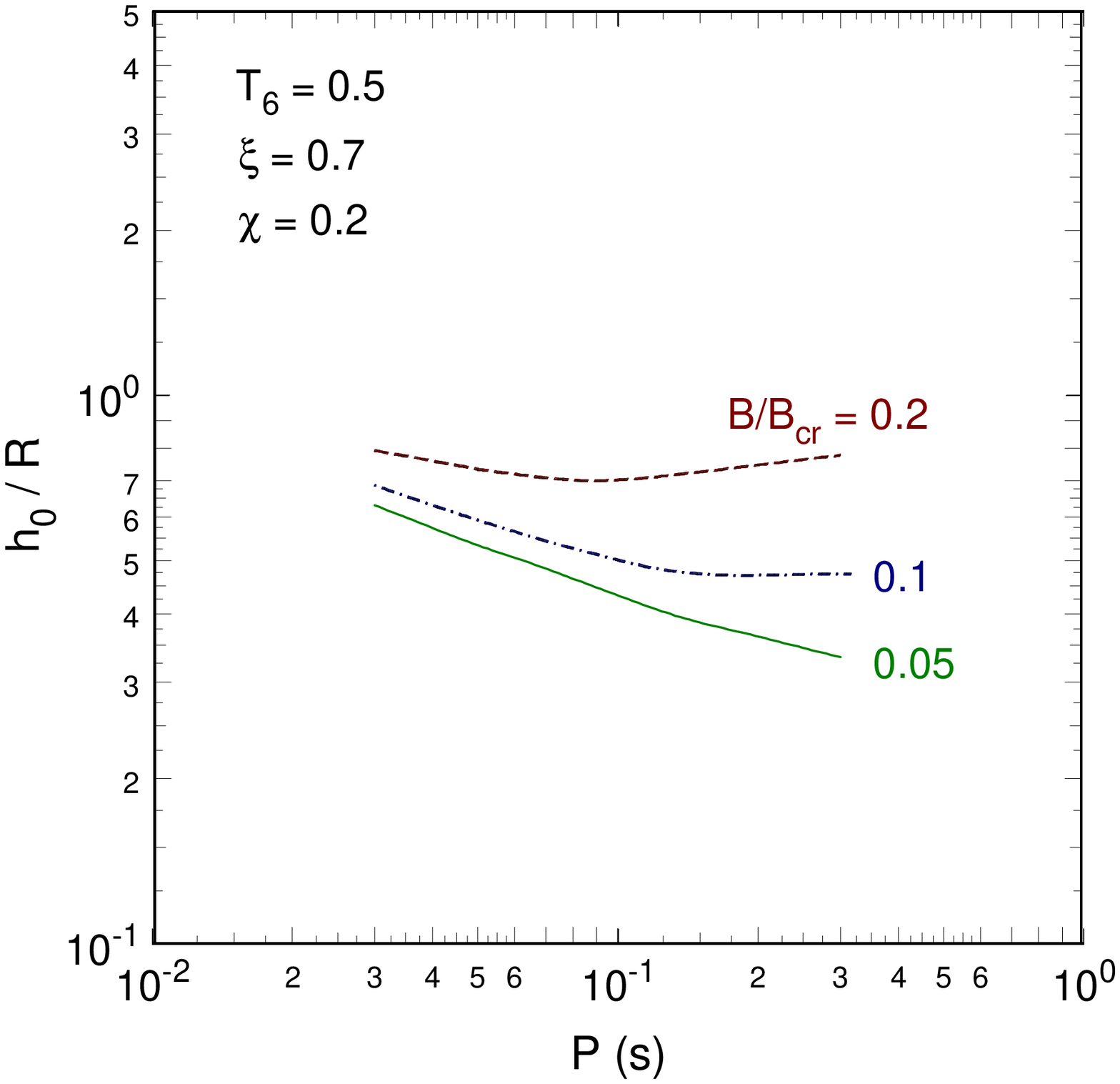}{0}{
Height at which electron energy losses, at Lorentz factors 
$\gamma^-_{\rm min}$, from CR and ICS are equal, above
which the electron PFF is controlled by CR, as a function of pulsar period and
surface value of magnetic field strength. 
   \label{fig:pffh0} }        

\figureout{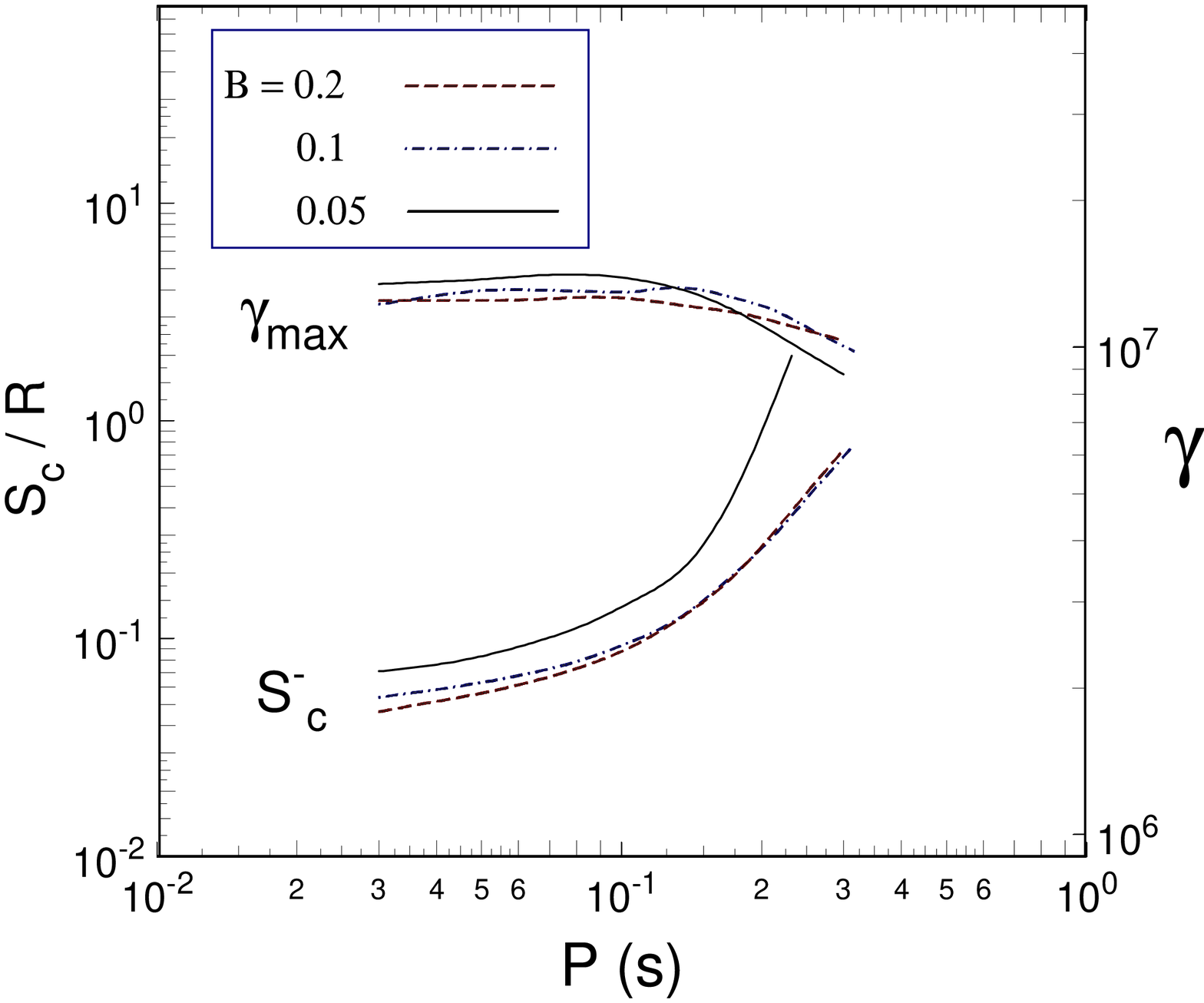}{0}{ 
Acceleration voltage (maximum electron Lorentz factor) $\gamma_{\rm max}$
and width, $S^-_c$, of the acceleration zone formed by a CR-controlled electron 
PFF at the height $h_0$ where control of the electron PFF switches from ICS to
CR (see Fig. \ref{fig:pffh0}), as a function of pulsar period and surface 
value of magnetic field strength.
   \label{fig:pffp} }        

\figureout{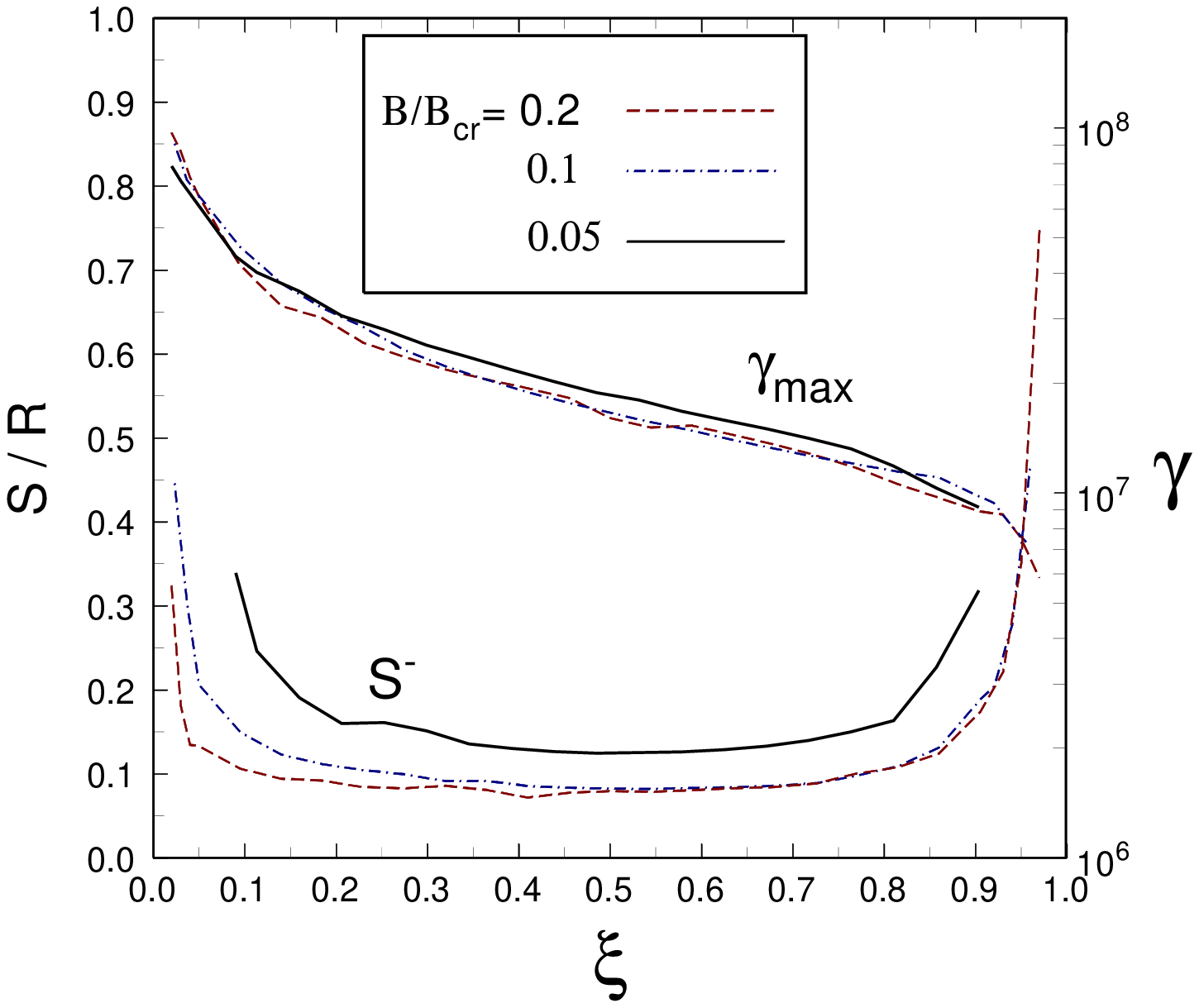}{0}{ 
Profiles of the acceleration voltage (maximum electron Lorentz factor) $\gamma_{\rm max}$
and width, $S^-_c$, of the acceleration zone formed by a CR-controlled PFF at the height 
$h_0$ where control of the electron PFF switches from ICS to CR (see Fig. \ref{fig:pffh0}), 
as a function of magnetic 
colatitude scaled to the polar cap half angle $\xi = \theta/\theta_0$, for different
values of surface magnetic field strength. 
   \label{fig:pffx} }        

\end{document}